\documentstyle[epsfig,setspace]{article}
\begin{document}
\title{Classical Vs Quantum Probability in Sequential Measurements}
\author{Charis Anastopoulos , \\
Department of Physics, University of Patras, 26500 Paras,
Greece\footnote{Email: anastop@physics.upatras.gr} }

\maketitle


\begin{abstract}
We demonstrate in this paper that the probabilities for sequential
measurements have features very different from those of single-time
measurements. First, they cannot be modeled by a classical
stochastic process. Second, they are contextual, namely they depend
strongly on the specific measurement scheme through which they are
determined. We construct Positive-Operator-Valued measures (POVM)
that provide such  probabilities. For observables with continuous
spectrum, the constructed POVMs depend strongly on the resolution of
the measurement device, a conclusion that persists even if we
consider a quantum mechanical measurement device or the presence of
an environment. We then examine the same issues in alternative
interpretations of quantum theory. We first show that multi-time
probabilities cannot be naturally defined in terms of a frequency
operator. We next prove that local hidden variable theories cannot
reproduce the predictions of quantum theory for sequential
measurements, even when the degrees of freedom of the measuring
apparatus are taken into account. Bohmian mechanics, however, does
not fall in this category. We finally examine an alternative
proposal that sequential measurements can be modeled by a process
that does not satisfy the Kolmogorov axioms of probability. This
removes contextuality without introducing non-locality, but implies
that the empirical probabilities cannot be always defined (the event
frequencies do not converge). We argue that the predictions of this
hypothesis are not ruled out by existing experimental results
(examining in particular the "which way" experiments); they are,
however, distinguishable in principle.
\end{abstract}

\renewcommand {\thesection}{\arabic{section}}
\renewcommand {\theequation}{\thesection. \arabic{equation}}
\let \ssection = \section \renewcommand{\section}{\setcounter{equation}{0} \ssection}

\section{Introduction}

\subsection{The main theme}

The topic of this paper is sequential quantum measurements and
their probabilistic description. We show that the construction of probabilities
for sequential measurements  is rather intricate and in some
aspects differs strongly from its analogues in classical
probability theory. For example, quantum multi-time probabilities
do not define a stochastic process. We can isolate specific points
of divergence between quantum and classical probability theory
(including hidden variable theories in the latter) and to argue
that these differences can be empirically determined, at least in principle.

The motivation for this line of inquiry is two-fold. First, the
determination of probabilities in sequential measurements is of
interest on its own right. It seems experimentally feasible, as it
is nowadays possible to construct sources that emit individual
systems. However, the construction of such probabilities from the
rules
 of standard
quantum theory is not as straightforward as it may seem, for the
relevant probabilities can not be obtained in a natural way from
the Hilbert space geometry. Assumptions about the physical
implementation of the measurement process are needed, and these
touch inevitably upon fundamental interpretational issues.

An immediate result of our analysis is that multi-time
probabilities are strongly dependent upon the specific
experimental set-up used in their determination.  For observables
corresponding to operators with discrete spectrum, one may
construct  a probability distribution rather simply. The same
procedure applied to observables with continuous spectrum leads to
probabilities that depend very strongly on an additional parameter
$\delta$. This parameter can be interpreted as the resolution of
the measurement device, but the dependence of the resulting
probabilities is so strong as to be highly counter-intuitive. This
dependence persists  even for samplings coarse-grained at a scale
much larger than $\delta$. An interesting corollary of this
analysis is that it is impossible to simulate by a stochastic
process the probabilities obtained from sequential measurements of
a quantum system.

The other motivation for this research is related to basic
interpretational issues of  quantum theory. Probabilities are
introduced in the quantum mechanical formalism through Born's
interpretation of the wave function. Born's rule is valid for
single-time measurement of one observable (or for a family of
compatible observables). In that case,  quantum theory is reduced
to a description in terms of classical probabilistic concepts,
which describe successfully the statistical outcomes of
experiments.

But once one moves away from this context, the coexistence between
quantum theory and classical probability theory becomes less
harmonious. This is highlighted by three representative theorems:
Bell's, Wigner's and Kochen-Specker's \cite{Bell64, KoSp67}.

The violation of Bell's inequalities (and their generalisations)
implies that local hidden variables theories are ruled out by
experiment. This may imply either quantum non-locality, or that it
is impossible to define a sample space for a physical system in
itself, without referring to the specific experiment that is
carried out. The latter property is referred to as {\em
contextuality} of quantum properties (or measurements). Wigner's
theorem is a representative of a more general result: it is not
possible to define a joint probability distribution for variables
that correspond to non-commuting operators. This can be argued to
be a form of contextuality, in the sense that there does not exist
a universal sample space to describe the outcomes of all possible
measurements that can be performed in an ensemble of quantum
systems. The Kochen-Specker theorem demonstrates a stronger form
of contextuality: it is impossible to assign definite values to a
physical observable without referring to the commuting set that is
measured along with it.

While all three theorems above suggest that quantum mechanical
properties (and consequently probabilities) are contextual, they do
not easily relate to empirical evidence. The observed violation of
Bell's inequalities may be attributed to non-locality rather than
contextuality, the measurement of incompatible observables involves
distinct experimental situations, whose outcomes cannot be
immediately compared, while the Kochen-Specker theorem refers to
idealized values of observables that a physical system possesses
prior to measurement (hence empirically inaccessible).

Sequential measurements on the other hand provide a ground, on
which the idea of contextuality can be explicitly tested. The
application of the  rules of standard quantum theory suggests that
two different measurement schemes will give rise to different
value for the probability of the same property of a physical
system, even if the initial state is assumed to be the same. Hence
the precise statistical study of the outcomes in sequential
measurements may in principle reveal unambiguously the contextual
character of quantum probability.

The problem is that we obtain much more contextuality than we
bargained for. Not only are multi-time probabilities dependent on
the measurement scheme through which they are determined, but they
seem to depend strongly on rather trivial details of the
measurement device. This is unavoidable, at least if we do not
abandon the usual rules of quantum theory. It is then questionable
whether it is possible to properly define a statistical ensemble
for sequential measurements, or even if any physical information
can be extracted from them.

This rather disturbing feature of multi-time probabilities
provides the motivation to seek an alternative account. We first
consider hidden variable theories. We prove that any local hidden
variable theory (deterministic or stochastic) that reproduces the
single-time probabilities of quantum theory cannot reproduce those
for multi-time probabilities. The only way to do so is by assuming
a non-local interaction between system and measuring device,
similar to the one appearing in Bohmian mechanics.

The other alternative we examine here is related to proposals
\cite{Ana01a, Ana03a}--see also \cite{Sor9497}--that it might be
possible to avoid contextuality (the constraints of Bell's and
Kochen-Specker's theorem) by assuming that quantum theory is
described by a "probability" measure that does not satisfy the
Kolmogorov axioms--in particular the additivity property. While a
non-additive measure is mathematically natural in multi-time
probabilities, its physical interpretation is somewhat
problematic. A non-additive measure cannot be interpreted in terms
of any empirical probabilities, which are obtained by the limit of
event frequencies. It only make sense if one assumes that the
event frequencies for sequential measurements do not converge to
probabilities. We explore further this idea, showing that it is
consistent with usual treatment of probabilities in quantum
theory, that it is natural from an operational point of view and
that it is in principle distinguishable from any alternative that
assumes that empirical probabilities for sequential measurements
always exist.

\subsection{The structure of this paper}
The paper is structured as follows.

In Section 2 we briefly review classical and quantum probability
theory, in order to set-up our conventions. We also provide some
preliminary mathematical arguments about the inequivalence between
the classical and quantum descriptions of sequential measurements.

Section 3 contains the central results of this paper. First, we
motivate the discussion on probabilities of sequential
measurements, focusing in particular on the fact that the quantum
mechanical correlation functions are complex-valued and have no
immediate correspondence in terms of objects that can be
immediately determined. Then we demonstrate that quantum logic
cannot be expected to hold in sequential measurements. This is
unlike single-time measurements for which the spectral theorem
together with Born's rule  guarantee that different measurement
schemes lead to the same probability assignment (assuming
identical preparation). Multi-time probabilities are therefore
highly contextual. We then discuss the description of multi-time
probabilities via Positive-Operator-Valued-Measures (POVMs). We
prove two theorems that demonstrate that it is not possible to
construct POVMs for sequential measurements compatible with the
single-time predictions of quantum theory. These results provide a
general proof of an often quoted statement that quantum mechanical
probabilities cannot be simulated by stochastic processes. We then
demonstrate different ways of constructing POVMs for a specific
class of multi-time measurements of position. These POVMs exhibit
a very strong dependence on properties of the measurement device
(its resolution) that persist even in highly coarse-grained
samplings. Finally we show that neither the consideration of a
fully quantum measuring device or of decoherence due to the
environment affect significantly these conclusions.

In section 4 we discuss other interpretational schemes, most
notably hidden variable theories and we demonstrate that the
predictions of quantum theory for sequential measurements are not
compatible with the assumption of local interactions between
measured system and measuring device. Finally in Section 5 we
consider the alternative proposal that probabilities for
sequential measurements cannot be defined because the relative
frequencies do not converge. The motivation for this proposal is
analysed in detail. We then demonstrate that it is compatible with
the predictions of single time quantum theory, that it is not
contradicted by some well-established results and that it is
possible to distinguish it unambiguously even in very simple
experimental set-ups.

\section{Classical Vs quantum probability}

\subsection{Basic facts}
We briefly describe here  the mathematical structure of classical
and quantum probability, in order to set-up our notations,
conventions and terminology for later use.

\subsubsection{Classical probability theory}
In classical probability one assumes that  all possible elementary
 alternatives lie in a space $\Omega$, the
{\em sample space}. Observables are functions on $\Omega$, and are
usually called {\em random variables}.
 The outcome of  any measurement can be phrased as a statement that the system is found in
a given subset $C$ of $\Omega$. Hence the set of certain
well-behaved (measurable)  subsets of $\Omega$ is identified with
the set of all coarse-grained alternatives of the system. To each
subset $C$,
 there corresponds an observable $\chi_C(x)$, the characteristic function of the set $C$.
It is defined as $\chi_C(x) = 1$ if $x \in C$ and $\chi_C(x) = 0$
otherwise. It is customary to denote the characteristic function
of $\Omega$ as $1$ and of the empty set as $0$.

If an observable $f$ takes values $f_i$ in subsets $C_i$ of
$\Omega$
\begin{equation}
f(x) = \sum_i f_i \chi_{C_i}(x) \label{clspth}
\end{equation}
A {\em state} is  intuitively thought of as  a preparation of a
system. Mathematically it is represented by a measure on $\Omega$,
i.e a map that assigns to each  alternative $C$ a
probability $p(C)$. A probability measure satisfies the Kolmogorov
conditions
\\ \\
- for all subsets $C$ of $\Omega$, $0 \leq p(C) \leq 1$ \\
- $p(0) = 0 ; p(1) = 1$. \\
- for all disjoint subsets $C$ and $D$ of $\omega$, $p(C \cup D) = p(C) + p(D)$ \\
\\
Due to (\ref{clspth}) one can define $p(f) = \sum_i f_i p(C_i)$;
$p(f)$ is  the mean value of $f$. In the case that $\Omega$ is a
subset of ${\bf R}^n$, the probability measures are defined in terms
of a probability distribution, i.e. a positive function on $\Omega$,
which we shall  denote as $p(x)$.
\begin{equation}
p(f) = \int dx p(x) f(x)
\end{equation}
\subsubsection{Quantum probability theory}
 The formalism of quantum mechanics incorporates
probability through Born's rule, which in its initial form asserts
that the square modulus $|\psi(x)|^2$ of Schr\"odinger's wave
function can be interpreted as a probability density for the
particle's position. In the abstract Hilbert space formulation
Born's interpretation can be implemented through the {\em spectral
theorem}: under rather general conditions we may assign a
Projection-Valued-Measure (PVM) $dE(\lambda)$ to each self-adjoint
operator $\hat{A}$. The PVM is  a map assigning to each measurable
set $U$ of $\hat{A}$'s spectrum $\sigma(\hat{A})$ a projection
operator $\hat{E}(U) = \int_U d\hat{E}(\lambda)  $, such that
$\hat{E}(U) = \chi_U(\hat{A})$, where $\chi_U$ is the
characteristic function of $U$. The projectors in the range of the
PVM reflect the Boolean algebra of the subsets of $\sigma(\hat{A}$
in the sense that
\\ \\
-- $\hat{E}(\emptyset) = 0, \; \; \hat{E}(\sigma(\hat{A})) =
\hat{1},$ \\
-- $ \hat{E}(U \cup V) = \hat{E}(U) + \hat{E}(V), \; U \cap V =
\emptyset, $ \\
-- $\hat{E}(U \cap V) = \hat{E}(U) \hat{E}(V).$
\\ \\
The spectral theorem implies that the Hilbert space $H$ is
isomorphic to that of square-integrable functions over
$\sigma(\hat{A})$, and as such the Born rule may be directly
applied: the probability for an event corresponding to $U \subset
\sigma(\hat{A}) $ is
\begin{eqnarray}
p(U) = Tr \hat{\rho} \hat{E}(U).
\end{eqnarray}
Given that $\hat{A} = \int  \lambda \, d\hat{E}(\lambda)$ the
standard relation between probabilities and expectation values can
be established.

It follows that for single-time measurements of a single
observable (or of many observables represented by mutually
commuting operators) quantum theory via the Born rule is
completely equivalent to classical probability theory.

The Copenhagen interpretation employs the formalism of quantum
theory to account for the outcomes of specific experiments. It
presupposes a split between the measured system, which is fully
quantum, and the measuring apparatus, which is part of the
classical world. While this creates the key problem of explaining
the classical description of an object that consists of
fundamentally quantum entities, it is fully self-consistent  at an
operational level, namely if we only employ quantum theory to
account for the statistics of measurement outcomes.

We shall adopt an operational stance in most discussions in this
paper. The reason for this choice is that the operational
description  is a core of quantum theory that refers immediately
to the concrete experimental situations, and the remarkable
success of quantum theory implies that  all contending
interpretation must accept it, either as a fundamental or as an
emergent theory.
 Still, we shall find it necessary in the
course of the argument to move beyond the operational description
and consider quantum measurement theory, namely the assumption
that the measuring apparatus is fully or partly quantum
mechanical.

\subsubsection{Probabilities and event frequencies}
 To apply a specific version of probability theory in a concrete
physical system,  one needs to be able to relate the numbers
obtained by the mathematical formalism to the  concrete
experimental data. This relation is achieved by the correspondence
of probability to relative frequencies of events in statistical
ensembles.  While it can be argued that relative frequencies do
not exhaust the physical content of probability theory, that the
latter can be interpreted in a way that refers to  individual
systems and not only statistical ensembles, and even that a {\em
definition} of probabilities from frequencies is highly
problematic,  any sharp quantitative test of a probabilistic
theory involves a comparison of theoretical probabilities to
empirical probabilities, which are obtained from event
frequencies.

Suppose for simplicity that the sample space of our system $\Omega
= {\bf R}$. We assume an experiment that
determines a value for $x \in {\bf R}$. Repeating the experiment
$n$ times we obtain a sequence $x_i, i = 1, \ldots, n$ of measured
values. We may then consider the relative frequency for the
proposition that the variable $x$ took value in the subset $U
\subset {\bf R}$. If $\chi_U$ is the characteristic function of
the set $U$\footnote{We assume that $U$ is a sufficiently
well-behaved set (like an open set) so that there is no
operational problem in ascertaining that $x \in U$.}, we define
the relative frequency for the occurrence of an event in $U$ for
the first $n$ experimental runs
\begin{eqnarray}
\nu_n(U) = \frac{1}{n} \sum_{i = 1}^n \chi_U(x_i).
\end{eqnarray}

The probability $p(U)$ associated to the event $U$ is the limit
\begin{eqnarray}
p(U) = \lim_{n \rightarrow \infty} \nu_n(U),
\end{eqnarray}
assuming of course that it exists.

 Since any actual determination of probabilities involves
 a finite number of runs, we can never establish the
 convergence of  frequencies. If, however, the description
 of the physical systems in terms of probabilities is valid, one
  expects that the relative rate of convergence
$\epsilon_n = \frac{| \nu_n(U) - p(U)|}{p(U)} \sim
\frac{1}{\sqrt{n}}$ by virtue of the central limit theorem. Hence
the fall-off of $\epsilon_n$ for large $n$ as $n^{-1/2}$ is a good
indication of convergence for the relative frequencies.

The mean value of a random variable $f(x)$ is similarly identified
as
\begin{eqnarray}
\langle f \rangle = \lim_{n \rightarrow \infty} \frac{1}{n}
\sum_{i = 1}^n f(x_i).
\end{eqnarray}

\subsection{The analogue of stochastic processes}

We saw that for the measurements of a single observable at a
single moment of time the predictions of quantum theory are fully
compatible with those of classical probability. It is then natural
inquire whether this correspondence passes through when one
considers the values of a single observable at more than one
moments of time.

In multi-time measurements the law of time evolution enters
explicitly. Let us assume a classical probabilistic system with
sample space $\Omega ={\bf R}$, whose probability density evolves
in time according to
\begin{eqnarray}
\frac{\partial}{\partial t} \rho = {\cal L} \rho,
\end{eqnarray}
where ${\cal L}$ is a positive, norm-preserving operator. The
formal solution of this equation is
\begin{eqnarray}
\rho_t = e^{{\cal L}t} \rho_0,
\end{eqnarray}
which can be written in terms of the integral kernel $g_t(x,x')$
of $e^{{\cal L}t}$
\begin{eqnarray}
\rho_t(x) = \int dx' g_t(x,x') \rho_0(x').
\end{eqnarray}

One may then define a probability measure $d \mu[x(\cdot)]$ on the
space of paths on $\Omega$  as a suitable limit of the expression
\begin{eqnarray}
d \mu(x_{t_1}, x_{t_2}, \ldots, x_{t_n}) = \rho_0(x_0)
g_{t_1}(x_0,x_1) g_{t_2 - t_1}(x_1, x_2) \ldots \nonumber
\\ g_{t_n - t_{n-1}}(x_{n-1}, x_n), dx_0 dx_1 dx_2 \ldots dx_n,
\label{prob}
\end{eqnarray}
which is defined on discrete-time paths.

The  reason it is possible to extend the single-time probability
to a stochastic probability measure is that the evolution law is
linear with respect to the probability density. In quantum theory
this is not the case; the evolution law is linear with respect to
the wave function and not the probability density. Hamiltonian
evolution mixes the diagonal elements of the density matrix (which
correspond to probabilities) with the off-diagonal ones (which
have no such interpretation).

It seems therefore not straightforward (if at all possible) to
extend the single-time probabilistic description of quantum theory
to a stochastic process. This conclusion will be verified by a
more rigorous analysis in section 3.3. Nonetheless, we can write
stochastic processes that reproduce some of quantum theory's
predictions. One such example is Nelson's stochastic mechanics
\cite{Nel66, Nel85}, which introduces a stochastic differential
equation on configuration space that can reproduce the expectation
values of the position observable at every moment of time.

\subsection{The history formalism}

The underlying reason that quantum evolution cannot be described
by a stochastic process is that the mathematically natural measure
on histories (or paths) does not satisfy the Kolmogorov axioms of
probability theory. This is particularly highlighted in the
consistent histories approach to quantum theory \cite{Gri84,
Omn8894, GeHa9093, Har93a}.

The basic object of this formalism is a history, namely  a {\em
time-ordered}  sequence of projection operators $\hat{P}_{t_1},
\ldots, \hat{P}_{t_n}$, and it corresponds to a time-ordered
sequence of propositions about the physical system. The indices
$t_1, \ldots, t_n$ {\it refer to the time a proposition is
asserted and have no dynamical meaning.} Dynamics are related to
the Hamiltonian $\hat{H}$, which defines the one-parameter group
of unitary operators $\hat{U}(s) = e^{-i\hat{H}s}$. In the
consistent histories approach a history is thought to correspond
to propositions about the physical system, not necessarily
associated to acts of measurement. Consistent histories is a
generalisation of Copenhagen quantum theory aiming to provide a
quantum mechanical description of individual systems.

The quantum rule for  conditional probability is that if the
property corresponding to the projector $\hat{P}_1$ is realized then
we may encode the information obtained in a change of the density
matrix\footnote{We shall argue later that this rule cannot be
applied freely, at least as far as measurements are concerned.}
\begin{eqnarray}
\hat{\rho} \rightarrow \frac{\hat{P}_1 \hat{\rho}
\hat{P}_1}{Tr(\hat{\rho} \hat{P}_1)}, \label{reduction}
\end{eqnarray}
hence the conditional probability the $\hat{P}_2$ will be realized
at time $t_2$ given that $\hat{P}_1$ was realized at $t_1$ equals
\begin{eqnarray}
\frac{ Tr \left(\hat{P}_2e^{-i \hat{H}(t_2-t_1)}\hat{P}_1 e^{-i
\hat{H}t_1}\hat{\rho} e^{i \hat{H}t_1} \hat{P}_1 e^{i \hat{H}t_1}
\hat{P}_2 e^{i \hat{H}(t_2-t_1)}\right)}{Tr(e^{-i
\hat{H}t_1}\hat{\rho} e^{i \hat{H}t_1} \hat{P}_1)},
\end{eqnarray}
leading to a probability for the joint realisation of $\hat{P}_1$
at $t_1$ and $\hat{P}_2$ at $t_2$
\begin{eqnarray}
Tr \left(\hat{P}_2e^{-i \hat{H}(t_2-t_1)}\hat{P}_1 e^{-i
\hat{H}t_1}\hat{\rho} e^{i \hat{H}t_1} \hat{P}_1 e^{i \hat{H}t_1}
\hat{P}_2 e^{i \hat{H}(t_2 - t_1) }\right)
\end{eqnarray}

For a general $n$-time history $\alpha = \{\hat{P}_{t_1},
\hat{P}_{t_2}, \ldots, \hat{P}_{t_n} \}$ this results generalizes a
s follows.  We define the class operator $\hat{C}_{\alpha}$ defined
by
\begin{equation}
\hat{C}_{\alpha} = \hat{U}^{\dagger}(t_n) \hat{P}_{t_n}
\hat{U}(t_n) \ldots \hat{U}^{\dagger}(t_1) \hat{P}_{t_1}
\hat{U}(t_1),
\end{equation}
which leads to a probability measure
\begin{eqnarray}
p(\alpha) = Tr \left( \hat{C}_{\alpha} \hat{\rho}
\hat{C}^{\dagger}_{\alpha} \right). \label{PROB}
\end{eqnarray}
These probabilities  do not define a genuine measure on the space of
histories. To see this, we consider two histories $\alpha =
\{\hat{P}_{t_1}, \hat{P}_{t_2}, \ldots, \hat{P}_{t_n} \}$ and $\beta
= \{\hat{P}'_{t_1}, \hat{P}_{t_2}, \ldots, \hat{P}_{t_n} \}$, such
that $\hat{P}_{t_1} \hat{P}'_{t_1} = 0$, the history $\{
\hat{P}_{t_1} + \hat{P}'_{t_1}, \hat{P}_{t_2}, \ldots, \hat{P}_{t_n}
\}$ is the logical join $\alpha \vee \beta$ of the histories
$\alpha$ and $\beta$. The probabilities, however, do not satisfy the
additivity condition
\begin{eqnarray}
p(\alpha \vee \beta) = p(\alpha) + p(\beta).
\end{eqnarray}

If the histories are interpreted as referring to measurements the
failure of the additivity condition is not (at first sight) a
problem, because each history corresponds to a different sequence of
YES-NO experiment and there is no {\em a priori} reason, why all
different experiments should be modeled by a common probability
measure. However, if histories are thought to correspond to
properties of individual system then the lack of a probability
measure becomes a problem.

In the consistent histories approach this is taken into account as
follows. We define  the decoherence functional as a complex-valued
function of pairs of histories: i.e. a map $d: {\cal V} \times
{\cal V} \rightarrow {\bf C}$. For two histories $\alpha$ and
$\alpha'$ it is  given by
\begin{equation}
d(\alpha, \alpha') = Tr \left( \hat{C}_{\alpha} \hat{\rho}_0
\hat{C}_{\alpha'}^{\dagger} \right). \label{decfun}
\end{equation}
The consistent histories  interpretation of this object is that
when $d(\alpha, \alpha') = 0$ for $\alpha \neq \alpha'$ in an
exhaustive and exclusive set of histories \footnote{ By exhaustive
we mean that at each moment of time $t_i$,
$\sum_{\hat{\alpha}_{t_i}} \hat{\alpha}_{t_i} = 1 $ and by
exclusive that $\hat{\alpha}_{t_i}  \hat{\beta}_{t_i} =
\delta_{\alpha \beta}$. Note that by $\alpha$ we denote  the
proposition with  the corresponding projector written as
$\hat{\alpha}$ with a hat.}, then one may assign a probability
distribution to this set as $p( \alpha) = d(\alpha, \alpha)$. The
value of $d(\alpha, \beta)$ is, therefore, a measure of the degree
of interference between the  histories $\alpha$ and $\beta$.

We end this section with a remark. We shall employ many
mathematical objects appearing in the consistent histories
approach throughout this paper(without a change in name or
notation). The reader should keep in mind that the focus of this
paper is the description of measurement outcomes through the rules
of standard quantum theory, hence the context and interpretation
of these objects are different from those in consistent histories.

\section{Sequential measurements in standard quantum theory}

\subsection{Multi-time correlation functions}

In classical probability theory there is no conceptual distinction
between single-time and multi-time measurements of a physical
system. If the sample space for the single-time measurement is
$\Omega$,  the sample space for $n$-time measurements as a
Cartesian product $\times_n \Omega_n$: the outcome of $n$
measurements of an observable $x$ is an ordered $n$-tuple of
values of $x$. In general, one may define an sample space
$\Omega^T$ of all paths from a time interval $T = [0, t]$ to
$\Omega$ and a corresponding stochastic measure
 $d \mu[x(\cdot)]$.
  One then immediately transfers the interpretation of
probabilities in terms of relative frequencies and  reconstructs the
statistical behavior of any observable on the multi-time sample
space.

The probabilities for the measurements of an observable $x$ is
most conveniently incorporated in the (unequal-time) correlation
functions of an observable $f(x)$,
\begin{eqnarray}
\langle f_{t_1} f_{t_2} \ldots f_{t_n} \rangle = \int d \mu
[x(\cdot)] F_{t_1}[x(\cdot)] F_{t_2}[x(\cdot)] \ldots
F_{t_n}[x(\cdot)],
\end{eqnarray}
in terms of the functions $F$ on $\Omega^T$ defined by
\begin{eqnarray}
F_t[x(\cdot)] = f(x(t)).
\end{eqnarray}

From an operational point of view, there is no problem in measuring
multi-time probabilities or correlation functions, as long as the
corresponding single-time measurements do not destroy the physical
system. The same is true for quantum mechanical systems: we may
consider for example a succession of Stern-Gerlach devices, or
microscopic particles leaving their trace in sharply localized
layers of recording material (we shall elaborate on such experiments
later). We therefore expect that quantum mechanics should allow us
to determine the values of the correlation functions, which can be
unambiguously determined from experiment.

The objects we usually call correlation functions in quantum
theory are expectation values of products of operators, such that
\begin{eqnarray}
\langle x_{t_1} x_{t_2} \rangle = \langle \psi|e^{i \hat{H}t_1}
\hat{x} e^{i \hat{H}(t_2-t_1)} \hat{x} e^{- i \hat{H}t_2}| \psi
\rangle. \label{cf}
\end{eqnarray}
These "correlation functions" are in general complex-valued, and
for this reason they have no interpretation in terms of the
statistics of measurement outcomes. Clearly, the construction and
interpretation of multi-time quantum probabilities involves many
more subtleties than their analogue in the single-time case.

In light of the discussion above, there are two questions that must be raised.\\ \\
--First, what is the physical meaning of the mathematically
natural
complex-valued correlation functions? \\
--Second, how can we employ the standard quantum mechanical
formalism (or slight generalisations thereof) to construct
real-valued correlation functions that would describe the
statistics of multi-time measurements?
\\ \\
Before proceeding to address these questions let us comment on a
rather naive answer that can be given to the second one: the
physically relevant correlation functions can be obtained by
elementary algebraic manipulations on the complex-valued ones,
taking for example their real part, or their totally symmetrizes
version etc. The immediate objection is that any such choice is
completely {\em ad hoc} with no justification in terms of the usual
principles of quantum theory. Why should we choose the real part of
the correlation function, rather than the imaginary part, or their
modulus? But even if we decide by fiat that a specific answer is the
correct one, the problem persists at the level of the probabilistic
interpretations: correlation functions must be related to
probabilities for sequential measurements. Hence any determination
of correlation functions must deal with the problem that the
mathematically natural probability measure for histories is
non-additive.

\subsection{Sequential measurements and quantum logic}
We now examine the definition of probabilities for multi-time
measurements. There is a substantial literature on this topic--see
for example \cite{ABL64,DaLe71, Davies, AAD85, Cave86, MiSu77,
BCL90, Hal93, SN01, Hol01}; indeed discussion of this issue can be
traced back to the early days of quantum mechanics. Our presentation
here aims to highlight the specific quantum mechanical features
through comparison with analogous 'experiments' in classical
probability.

For ideal measurements, one may employ equation (\ref{PROB}) for the
probabilities. This expression defines a non-additive measure on the
space of histories (we use the word histories heuristically here to
denote a temporal succession of measurement outcomes). On the other
hand, any empirical probability that is constructed by event
frequencies should satisfy the additivity condition. This is an
apparent contradiction.

 As a first step towards an answer we shall elaborate on
specific features of single-time measurements. In any well-designed
experiment, we need to
 guarantee that the results do not depend too strongly on specific details the  measurement device. The reasons for
 that are epistemological  (experiments must be
 reproducible) but also practical: minor details of the
 measurement device should not affect the
 experimental outcomes significantly. They should ideally be hidden within the
 sampling or  systematic errors of the experiment. Moreover, it
 would be highly desirable if different measurement schemes for
 the same observable and with the same preparation procedure
 should give compatible (if not  identical) results.

 We may consider for example two different measurement schemes for
 the position of a particle. In the first, we  assume  a source
 emitting electrons with well defined momentum in the z-direction,
 but with significant spread in the x and y directions \footnote{For example,
 the $z$-degrees of freedom may be represented by the
wave function $\psi(z) = \frac{1}{(2 \pi \sigma_z)^{1/4}} e^{-
\frac{z^2}{4 \sigma_z^2} + i p_z z}$, such that the spread $\Delta
p_z = 1/\sigma_z << p_z$. This set-up corresponds to a measurement
at a reasonably well-specified moment of time. }. At a
 specific distance from the source we place a photographic plate that records the electron's position.
  This set-up is equivalent to a single-time
 measurement of the electron's $x$ and $y$ coordinates. The
 distribution of electrons on the screen corresponds to a
 probability distribution, which modulo sampling errors is given by Born's
 rule. The number of electrons found in a subset $U$ of the plate is  proportional to
  $p(U) = \int_U dx dy |\psi(x,y)|^2$.

 We may also consider a filter measurement of the electron's
 position, by placing instead of a photographic plate a curtain
 with a hole corresponding to the subset $U$. Any detector placed
 behind the whole will  register a number of particles proportional
 to $Tr \rho \hat{P}(U)$. This type of measurement is known as a
 {\em YES-NO experiment}, because it can only admit two answers: the
 particle passing through $U$ or not.

 The important point is that the value for the probability $p(U)$ in the
 experiment with the photographic plate  coincides with
 that obtained from the YES-NO experiment. Moreover, if we carry a
 sufficiently large number of YES-NO experiments differing only in
 the position of the hole, we will obtain sufficient information to
 fully reconstruct the probability distribution of the first
 experiment. In other words, in single-time quantum theory, YES-NO
experiments contain the full probabilistic information about a
quantum system. The empirical probabilities for a sample set $U$
are the same in all measurement schemes that correspond to the
same preparation of the physical system, modulo sampling and
systematic errors. This universality is often referred to as
defining a logic for quantum measurements, quantum logic. It is in
effect a consequence of the {\em spectral theorem}.

      \setlength{\unitlength}{0.14cm}
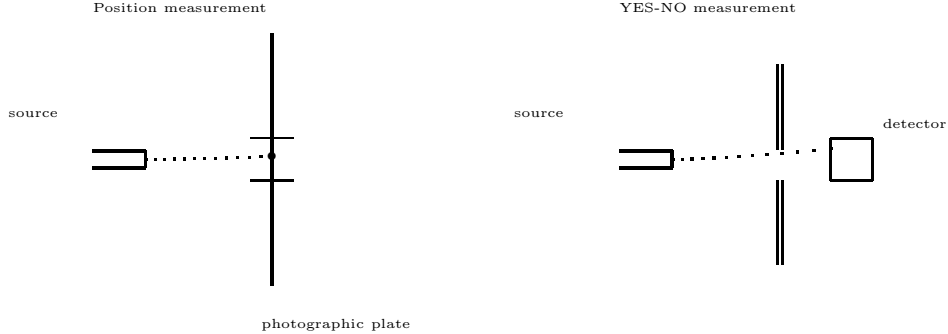
\begin{figure}
\begin{picture}(36,00)(-12,20)
\put (0, 14) {\tiny Position measurement}

 \put (-8, 4){{\tiny source}} \thicklines \put (0, 0.8)
{\line(1,0){5}} \put (0, -0.8){\line(1,0){5}} \put
(5,-0.8){\line(0,1){1.6}}

 \put(17,12){\line(0,-1){24}}
\put(17,0.3){\circle*{0.7}}

\qbezier[12](5,0), (10, 0), (17,0.3) \thicklines \put (50, 0.8)
{\line(1,0){5}} \put (50, -0.8){\line(1,0){5}} \put
(55,-0.8){\line(0,1){1.6}}

\put(65,1){\line(0,1){8}} \put(65.5, 1){\line(0,1){8}}
\put(65,-2){\line(0,-1){8}} \put(65.5,-2){\line(0,-1){8}}

\put (40, 4){{\tiny source}} \put (16, -16) {{\tiny photographic
plate}}

 \put(70,2){\line(1,0){4}}
\put(74,2){\line(0,-1){4}} \put(74,-2){\line(-1,0){4}}
\put(70,-2){\line(0,1){4}} \put(75,3){{\tiny detector}}

\qbezier[12](55, 0), (60, 0), (70,1)

\put (50, 14) {\tiny YES-NO measurement}

\thinlines \put (15, 2) {\line (1,0){4}} \put (15, -2) {\line
(1,0){4}}
\end{picture} \\ \\ \\ \\ \\ \\ \\ \\ \\ \\ \\
\caption{\scriptsize Single-time position measurement Vs single
time filter measurement of position.}
\end{figure}

We now return to the analysis of multi-time measurements. If the
only possible multi-time experiment that could be carried out were
of the YES-NO type, there would be no downright problem from the
non-additivity of (\ref{PROB}), at least not a worse problem than
appearing in any other quantum "paradox".

To see this one may consider the following two-time YES-NO
experiment measuring the position of a particle. We assume a
source of electrons prepared in the same state as in the previous
examples. At fixed distances from the source and parallel to the
x-y plane we place two curtains with holes corresponding to the
subsets $U_1$ and $U_2$ of the $x-y$ plane. Behind the second slit
we place a particle detector.

Repeating the experiments above $n$ times, we record the number of
times the detector click, thus constructing the sequence of
relative frequencies $\nu_n(U_1, t_1; U_2, t_2)$, and from it the
corresponding probability $p(U_1, t_1; U_2, t_2)$. To construct
the probability $p(U_1', t_1; U_2, t_2)$, for a different slit
corresponding to $U_1'$ we have to {\em change the experimental
configuration}, and similarly for $p(U_1 \cup U_1', t_1; U_2,
t_2)$. Hence the probabilities $p(U_1, t_1; U_2, t_2)$, $p(U_1',
t_1; U_2, t_2)$ and $p(U_1 \cup U_1', t_1; U_2, t_2)$ do not refer
to the same experimental set-up, and there is no contradiction
between Eq. (\ref{PROB}) and the additive character of relative
frequencies.

\begin{figure}
\begin{picture}(36,00)(-12,20)

 \thicklines \put (10, 0.8)
{\line(1,0){5}} \put (10, -0.8){\line(1,0){5}} \put
(15,-0.8){\line(0,1){1.6}}

\put(25,1){\line(0,1){8}} \put(25.5, 1){\line(0,1){8}}
\put(25,-2){\line(0,-1){8}} \put(25.5,-2){\line(0,-1){8}}

\put (0, 4){{\tiny source}}

 \put(40,2){\line(1,0){4}}
\put(44,2){\line(0,-1){4}} \put(44,-2){\line(-1,0){4}}
\put(40,-2){\line(0,1){4}} \put(45,3){{\tiny detector}}

\qbezier[12](15, 0), (25, 0), (40,1)

\put(35,3){\line(0,1){6}} \put(35.5, 3){\line(0,1){6}}
\put(35,-1){\line(0,-1){9}} \put(35.5,-1){\line(0,-1){9}}
\end{picture} \\ \\ \\ \\ \\ \\ \\ \\ \\ \\
\caption{\scriptsize A two-time YES-NO measurement of a particle's
positions.}
\end{figure}
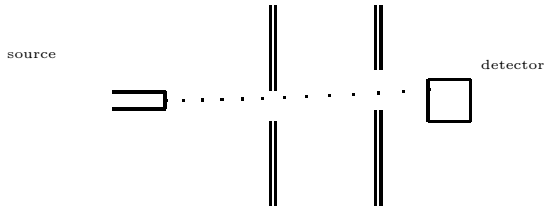

However, YES-NO measurements are not the only one possible in
practice. For measurements at a single moment of time they contain
all the probabilistic information of quantum theory, but this does
not hold for sequential measurements. To see this, let us consider
the following scheme for a  two-time measurement of position.  We
assume a particle source as before, which can be controlled so
finely as to emit a single particle at a time. Two thin sheets of
penetrable material are placed  one after the other in front of
the particle source, both parallel to the x-y plane. Particles
leave tracks as they cross through the sheets, and one may then
determine their
 $x$ and $y$ coordinates.

Each time the source emits a particle we record the readings
$(x_1,t_1;x_2,t_2)_n$;  $n$ labels the experimental runs and the
$y$ coordinate is suppressed for brevity. We thus construct a
sequence of measurement outcomes. From this one    defines the
sequence $\nu_n(U_1,t_2;U_2,t_2)$  for each pair of subsets $U_1$
of the sheet at $t_1$ and $U_2$ of the sheet at $t_2$. One obtains
the probability $p(U_1,t_1;U_n,t_n)$ as the limit
$\nu_n(U_1,t_1;U_n,t_n)$ as $n \rightarrow \infty$--assuming it
exists.
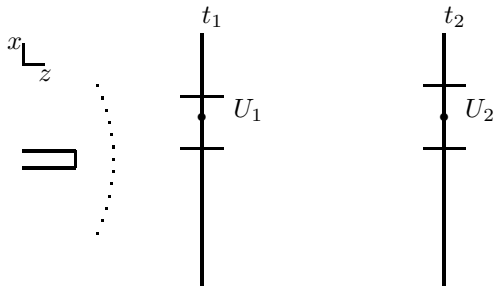
\begin{figure}
\begin{picture}(36,00)(-12,20)
 \thicklines \put (0, 0.8) {\line(1,0){5}} \put
(0, -0.8){\line(1,0){5}} \put (5,-0.8){\line(0,1){1.6}}
\put(40,12){\line(0,-1){24}}
 \put(17,12){\line(0,-1){24}}
\put(17,4){\circle*{0.7}} \put(40,4){\circle*{0.7}}
 \put (17, 13){$
t_1$} \put(40,13){$ t_2$}
\thicklines \put(0,9) {\line(1,0){2}} \put(1.5,7.5){$z$}
\put(0,9){\line(0,1){2}} \put(-1.5,10.5){$x$}
\qbezier[12](7,7), (10, 0), (7,-7)

\put (15, 6) {\line(1,0) {4}}  \put (15, 1) {\line(1,0) {4}} \put
(38, 7) {\line(1,0) {4}}  \put (38, 1) {\line(1,0) {4}}

\put(20,4){$U_1$} \put (42,4){$U_2$}
\end{picture}
\\ \\ \\ \\ \\ \\ \\ \\ \\ \\ \\ \\
\caption{\scriptsize A two-time measurement of a particle's
position. The particle leaves a trace on the plates of penetrable
material. One then samples the data into specific subsets of the
plates, in order to construct the corresponding multi-time
probabilities.}
\end{figure}

Unlike YES-NO experiment the sequences $\nu_n(U_1,t_1;U_2,t_2)$
constructed for different choices of the sample sets all refer to
the same experimental set-up. They should therefore satisfy the
additivity condition (modulo sampling and systematic errors)
\begin{eqnarray}
\nu_n(U_1,t_1;U_2,t_2) + \nu_n(U_1',t_1;U_2,t_2) = \nu_n(U_1 \cup
U_1' ,t_1;U_2,t_2),
\end{eqnarray}
since they refer to indivisible and specific measurement events.
It follows that the probabilities (\ref{PROB}) do not describe the
outcomes of this experiment. This conclusion holds for any
multiple-time measurement, in which any possible alternative of
the observable can be recorded at each moment of time, provided
that the corresponding operator does not commute with the
Hamiltonian. One could consider, for example, a succession of two
Stern-Gerlach apparatuses, with different directions of their
magnetic fields placed in such a position as to measure the spin
of the particle in the direction ${\bf n}$ at time $t_1$ and in
the direction ${\bf n}'$ at time $t_2$.

 Note also that this
thought-experiment presupposes that we record the trace of each
particle individually on the sheets. It is, therefore, essential
that in each individual run of the experiment the source emits only
a single particle. If we perform this experiment with beams of
particles, we will not have sufficient statistical information to
construct the two-time probabilities. We would not be able to
ascertain that the particle found recorded in $x_1$ at time $t_1$ is
the same with the particle recorded in $x_2$ at time $t_2$. The most
we could obtain would be the two marginal probability distributions
for the probability density at $t_1$ and the probability density at
$t_2$\footnote{Experiments like that of Fig. 2 involve only a single
act of detection. Hence, even if they are formally a two-time YES-NO
experiment they can also be described as a single-time measurement
of a system, whose wave function satisfies specific boundary
conditions on the walls. This is the reason we shall ignore them and
study exclusively measurement schemes similar to that in Fig. 3. }.

There is one point that needs to be highlighted in our discussion.
It is well accepted in quantum theory that the presence of an
intermediate measurement affects the state of the system and for
this reason the experimental outcomes depend strongly on the whether
an intermediate measurement has been carried out. However, the same
statement could be made for sequential measurements in a {\em
classical} probabilistic system. However, as we shall explicitly
prove in section 4,  classical probability cannot give rise to the
degree of contextuality inherent in quantum theory even if the
coupling to  measurement devices is taken into account. To
demonstrate this in detail, we need first to expand on the
construction of probabilities corresponding to the
thought-experiments of Fig. 3.

\subsection{POVMs and their applicability}

\subsubsection{POVMs and their properties}
Unlike single-time measurements, sequential measurements cannot be
described by the spectral projectors of a self-adjoint operator.
It is therefore necessary to employ a generalisation of the notion
of quantum mechanical observables, namely the
Positive-Operator-Valued Measures (POVMs).

  A POVM is map that assigns
to each measurable subset $U$ of a sample space $\Omega$  a
positive operator $\hat{\Pi}(U)$, such that \\
-- $\hat{\Pi}(\Omega)= \hat{1}, \; \; \hat{\Pi}(\emptyset) = 0$ \\
-- $ \hat{\Pi}(U \cup V) = \hat{\Pi}(U) + \hat{\Pi}(V), \; U \cap
V = \emptyset $.
\\ \\
A POVM can therefore define a probability density on $\Omega$ by
\begin{eqnarray}
p(U) = Tr \left( \hat{\rho}\hat{\Pi}(U) \right).
\end{eqnarray}
POVMs are  generalisations of PVMs, usually thought to correspond
to unsharp measurements. Indeed, if we denote by $\lambda$ the
points of the spectrum of a self-adjoint operator $\hat{A}$, we
may define a POVM as
\begin{eqnarray}
\hat{\Pi}(U) = \int d \lambda \, \chi_U^{\delta}(\lambda) \, |
\lambda \rangle \langle \lambda|, \label{POVMex}
\end{eqnarray}
in terms of a family of smeared characteristic functions
$\chi_U^{\delta}$. (For smeared characteristic functions and their
properties see appendix A).

For sufficiently coarse sets $U$, the positive operators
$\hat{\Pi}(U)$ are close to true projectors. One may estimate that
\begin{eqnarray}
|Tr\hat{\rho} (\hat{\Pi}(U) - \hat{\Pi}(U)^2)| \leq Tr
\hat{\rho}|\hat{\Pi}(U) - \hat{\Pi}(U)^2)| \nonumber \\
\leq \int
d \lambda |\chi_U^{\delta}(\lambda) -
[\chi_U^{\delta}(\lambda)]^2| < c \delta,
\end{eqnarray}
with $c$ a constant of order unity. Also for states $\rho$ with
spreads in $\hat{A}$ much larger than $\delta$ we may compute (see
the Appendix A)
\begin{eqnarray}
|Tr\hat{\rho} (\hat{\Pi}(U) - \hat{\Pi}(U)^2)| < c'
\frac{\delta}{L} Tr
 \hat{\rho} \hat{\Pi}(U), \label{ineq4}
\end{eqnarray}
where $L$ is the size of $U$.

When a POVM $\hat{\Pi}$ is defined on a sample space $\Omega =
\Omega_1 \times \Omega_2$, we denote the POVMs
$\hat{\Pi}(\Omega_1, \cdot)$ and $\hat{\Pi}(\cdot, \Omega_2)$,
defined on $\Omega_2$ and $\Omega_1$ respectively, as the {\em
marginal POVMs} of $\hat{\Pi}$.

\subsubsection{POVMs for sequential measurements: non-go theorems}

One possibility that should be first considered is that the
arguments leading to equation (\ref{PROB}) are somehow inadequate to
account for the multi-time experiment we considered earlier and that
a different procedure should allow us to define proper probabilities
for multi-time measurements.

The most general way to define a probability distribution that is
linear with respect to the density matrix is through POVMs. One
could therefore conjecture the existence of a POVM on the sample
space $\otimes_n\Omega_n$ for the n-time measurements. There are,
however, limitations \footnote{The results implied from propositions
1 and 2 seem to be well accepted in the consideration of sequential
measurements. Even though they are rather elementary, we are not
aware of any explicit proof in the literature, and for this reason
we include the proof in the text. They are essential for the
development of the arguments in Section 4.}
\\ \\
{\bf Proposition 1.} There exists no POVM for $n$-time
measurements of an observable $\hat{x}$ compatible with the
single-time predictions of quantum theory, unless $\hat{x}$
commutes with the system's Hamiltonian $\hat{H}$.
\\ \\
We consider without loss of generality a POVM for a two-time
measurement. We denote by $\Omega$ the spectrum of $\hat{x}$, and
by $\hat{P}(U)$ the spectral projectors of $\hat{x}$, $U \subset
\Omega$. The POVM $\hat{E}(\cdot,t_1; \cdot, t_2)$ assigns to each
pair of sample sets $U_1, U_2 \subset \Omega$ a positive operator
$\hat{E}(U_1,t_1; U_2, t_2)$. It should be compatible with the
single-time predictions of quantum theory, namely
\begin{eqnarray}
Tr \left( \hat{\rho}  \hat{E}(U_1, t_1; \Omega, t_2)  \right) = Tr
\left(\hat{\rho} e^{i \hat{H} t_1} \hat{P}(U_1) e^{-i \hat{H} t_1}
\right) \\ \nonumber Tr \left( \hat{\rho}  \hat{E}(\Omega, t_1;
U_2, t_2) \right) = Tr \left(\hat{\rho} e^{i \hat{H} t_2}
\hat{P}(U_2) e^{-i \hat{H} t_2} \right). \label{compatprob}
\end{eqnarray}
Since this should hold for all $\hat{\rho}$,  the marginals of the
POVM $\hat{E}$ are PVM's, namely
\begin{eqnarray}
\hat{E}(U_1, t_1; \Omega, t_2) = e^{i \hat{H} t_1} \hat{P}(U_1)
e^{-i
\hat{H} t_1} \label{marg1} \\
\hat{E}(\Omega, t_1; U_2, t_2) = e^{i \hat{H} t_2} \hat{P}(U_2)
e^{-i \hat{H} t_2}. \label{marg2}
\end{eqnarray}

There is a general result (see e.g. Theorem 2.1 of reference
\cite{Davies}) that  any POVM, whose marginals are PVMs, is itself
a PVM, it commutes with its marginals and can be written as the
marginals' product. Hence
\begin{eqnarray}
[e^{i \hat{H} t_1} \hat{P}(U_1) e^{-i \hat{H} t_1}, e^{i \hat{H}
t_2} \hat{P}(U_2) e^{-i \hat{H} t_2}] = 0,
\end{eqnarray}
 and since this property
holds for all $t_1, t_2$ and subsets $U_1, U_2$, it follows that
$[\hat{x}, \hat{H}] = 0$. The probability measure (\ref{PROB}) is
additive in that case and the correlation functions are real-valued.
It follows that in the generic case the probabilities for $n$-time
measurements cannot be modeled by a stochastic process, because the
latter can only be defined if a compatibility condition of the form
(\ref{compatprob}) is satisfied (see the discussion in section
3.5.1).

One, however, may object that the requirement that the single-time
marginals of the POVM's are  projectors is too stringent. The
physical set-up of a two-time measurement is different from that
of a single-time measurement, and there is no {\em a priori}
reason for the marginals of the POVM to reduce to those of the
single-time measurement. One cannot argue so much against equation
(\ref{marg1}). If $t_1 < t_2$ the measurement outcomes at $t_1$
should not depend on whether or not we choose to perform a second
measurement later. However, Eq. (\ref{marg2}) may very well be
problematic, because the physical system has already interacted
with a measuring device, while in the single-time measurement the
evolution has been purely unitary.

 Still, even this less restrictive case
(namely only equation (\ref{marg1}) being satisfied) leads to the
same conclusions. The proof involves only a few small changes from
the earlier one, but we reproduce it here for concreteness.

We  consider without loss of generality $t_1 = 0$, $t_2 = t$. For
the sample sets $U_1$, $U_2 = \Omega - U_1$, $V_1$, $V_2 = \Omega
- V_1$ we define the positive operators
\begin{eqnarray}
\hat{E}_{ij} &=& \hat{E}(U_i,0; V_j,t), \\
\hat{K}_i &=& \hat{E}(U_i,0;\Omega, t), \\
\hat{L}_i &=& \hat{E}(\Omega,0; V_i,t).
\end{eqnarray}
By assumption $\hat{K}_i$ is a projector, while $\hat{L}_i$ is a
general positive operator.

By definition $ 0 \leq \hat{E}_{ij} \leq \hat{K}_j$, for both
values of $i$ \footnote{ $\hat{A} \leq \hat{B}$ means that
$\hat{B} - \hat{A}$ is a positive operator.}. Since $\hat{K}_i$ is
a projector, $\hat{E}_{ij}$ lies in the closed linear  subspace
corresponding to $\hat{K}_j$, with every $j$ taken separately.
Hence $\hat{E}_{i1}$ commutes with $\hat{K}_1$ and $\hat{E}_{i2}$
commutes with $\hat{K}_2$. Since $\hat{K}_2 = \hat{1} -
\hat{K}_1$, also $[\hat{E}_{i1}, \hat{E}_{i2}] = 0$. Since
$\hat{L}_i = \hat{E}_{i1} + \hat{E}_{i2}$, the operators
$\hat{E}_{i1}, \hat{E}_{i2}$ also commute with $\hat{L}_i$. Again
by definition  $0 \leq \hat{E}_{ij} \leq \hat{L}_i$ and since
$\hat{E}_{ij}$ lies in the closed-linear subspace corresponding to
$\hat{K}_j$ we obtain $0 \leq \hat{E}_{ij} \leq \hat{K}_j
\hat{L}_i \hat{K}_j = \hat{K}_j \hat{L}_i$. Since $\hat{1}=
\sum_{ij} \hat{E}_{ij} \leq \sum_j \hat{K}_j \hat{L}_i \leq
\hat{1}$, we obtain that $\hat{E}_{ij} = \hat{L}_i \hat{K}_j$. We
therefore conclude
\\ \\
{\bf Proposition 2}. A POVM for sequential measurements satisfies
(\ref{marg1}), only if its  marginals commute.
\\ \\
This implies in particular that the marginal $\hat{E}(\Omega, t_1;
U_2, t_2)$ cannot be a POVM of type (\ref{POVMex}) corresponding to an unsharp
measurement of $\hat{x}$, unless $[\hat{x}, \hat{H}] = 0$.

The assumption that equation (\ref{marg1}) holds is valid for ideal
measurements, like for instance the ones corresponding to
measurements of observables with discrete spectrum. The
generalisation of this result for non-ideal measurements is
straightforward.

 We conclude that we cannot
construct POVMs that provide the probabilities for multi-time
measurements in quantum systems, if we require that they reproduce
faithfully (or even approximately) the predictions of single-time
quantum theory. This, however, does not imply that we cannot
construct any such probabilities in a way compatible with the
predictions of single-time quantum theory.  POVMs provide the most
general way to construct probability densities on a sample space as
a {\em linear map } of the quantum state $\hat{\rho}$. If we break
linearity (and hence assume that the resulting construction will not
respect
 the convexity properties of the space of states) such an
assignment may be possible. However, probabilities defined through
such a procedure cannot be obtained from a measure of the form
(\ref{prob}) (corresponding to a Markov process), because such a
measure would be linear with respect to the initial density matrix.
We shall take up this issue again in section 4.2.

\subsection{Constructing POVMs for sequential measurements}

Propositions 1 and 2 above demonstrate the degree of contextuality
in sequential quantum measurements. They do not imply, however, that
no POVMs exist that provide the probabilities of sequential
measurements. Indeed, probabilities for sequential measurements have
been considered extensively in the literature. We shall construct
such POVMs in detail, in order to demonstrate that they are not only
mathematically natural, but also physically reasonable.

\subsubsection{Ideal measurements}
We first consider the case of measuring an observable $\hat{x} =
\sum_i \lambda_i \hat{P}_i$ with discrete spectrum. Writing
$\hat{Q}_i = e^{i \hat{H}t} \hat{P}_i e^{-i \hat{H}t}$, we
construct the probabilities for the most-fine grained two-time
results
\begin{eqnarray}
p(i,0; j,t) = Tr (\hat{Q}_j \hat{P}_i \hat{\rho}_0 \hat{P}_i) =
\langle i |\hat{\rho}_0|i \rangle |\langle i|e^{-i \hat{H}t}| j
\rangle|^2 \label{finegrained}
\end{eqnarray}

Irrespective of the interpretation of the measurement process, the
probabilities (\ref{finegrained}) refer to the most elementary
alternatives that can be unambiguously determined in the
experimental set-up corresponding to the sequential measurement of
$\hat{x}$. Therefore, they can be  employed to construct
probabilities for general sample sets $U_1$, $U_2$ on the spectrum
$\Omega$ of $\hat{x}$, namely
\begin{eqnarray}
p(U_1,0; U_2, t) = \sum_{i \in U_1} \sum_{j \in U_2} p(i,0; j,t).
\label{POVM2}
\end{eqnarray}
The total probability is normalized
\begin{eqnarray}
p(\Omega,0; \Omega,t) = \sum_{ij}Tr (\hat{Q}_j \hat{P}_i
\hat{\rho}_0 \hat{P}_i) = 1.
\end{eqnarray}
Hence Eq. (\ref{POVM2}) defines a POVM for two-time measurements.

Note that as a result of the construction above, the probabilities
$p(U_1,0; U_2, t)$ for general samplings do not depend on $U_1$
and $U_2$ through the corresponding projectors $\hat{P}_{U_1}$ and
$\hat{P}_{U_2}$. This strengthens the conclusion of section 3.2
that there is no quantum logic interpretation for multi-time
measurements. In classical probability we use the same
mathematical object (a characteristic function for a subset of the
sample space) to represent both a concrete measurement outcome and
a statement about a measurement outcome. In multi-time quantum
measurements this is no longer the case: a coarse-grained
projector $\hat{P}_U$ cannot represent a proposition that the
outcome of the corresponding measurement lies within $U$: it can
only represent a genuine physical event \cite{Ana04a}.

\subsubsection{Continuous spectrum}

The situation is  more complex when one considers observables with
continuous spectrum, such as position. In that case there are no
fine-grained projectors and the choice of the elementary quantum
probabilities, from which one may build the general probabilities
for measurement outcomes cannot be made  uniquely. We shall see
that this implies that the probabilities are very strongly
dependent on minor properties of the measurement device, so
strongly in fact as to put into question whether the definition of a statistical
ensemble is practically meaningful.

The immediate generalisation of Eq. (\ref{finegrained}) for the
measurement of an operator with a continuous spectrum is
\begin{eqnarray}
p(x_1,0; x_2,t) =  |\langle x_1 |\hat{\rho}_0|x_1 \rangle|^2
|\langle x_1|e^{-i \hat{H}t}| x_2 \rangle|^2.
\end{eqnarray}
This, however, does not define a proper probability density, because
it is not normalized to unity
\begin{eqnarray}
\int dx_1 \int dx_2 p(x_1,0; x_2,t) = \infty.
\end{eqnarray}
This is due to the fact that there can be no measurements of
infinite accuracy. One has, therefore, to take into account the
finite width of any position measurement, say $\delta$. This
quantity depends on the properties of the measuring device--for
example the type of the material that records the particle's
position.

 The simplest procedure (but  not the most natural one)
is to consider the measurement of a self-adjoint operator
$\hat{x}_{\delta} = \sum_i x_i \hat{P}^{\delta}_i$, where
$\hat{P}^{\delta}_i$ is a projection operator corresponding to the
interval $[x_i - \frac{\delta}{2}, x_i + \frac{\delta}{2}]$. In
that case we may immediately construct the fine-grained
probabilities
\begin{eqnarray}
p_{\delta}(i,0; j,t) = Tr (\hat{Q}^{\delta}_j \hat{P}^{\delta}_i
\hat{\rho}_0 \hat{P}^{\delta}_i), \label{fine}
\end{eqnarray}
from which we may construct probabilities for general sample sets
$U_1$ and $U_2$:
\begin{eqnarray}
p_{\delta}(U_1,0;U_2,t) = \sum_{i \in U_1} \sum_{j \in U_2} Tr
(\hat{Q}^{\delta}_j \hat{P}^{\delta}_i \hat{\rho}_0
\hat{P}^{\delta}_i).
\end{eqnarray}
 Strictly speaking one may only consider sample sets that are
unions of the elementary sets that define our lattice. If,
however, the size of the sample sets $L$ is much larger than
$\delta$, we may approximate the summation with an integral. This
amounts to defining the continuous version of probabilities
(\ref{fine})
\begin{eqnarray}
p_{\delta}(x_1,t_1;x_2, t_2) = \hspace{4cm}\\ \nonumber Tr \left(
e^{i\hat{H}(t_2-t_1)} \hat{P}^{\delta}_{x_2}
e^{-i\hat{H}(t_2-t_1)} \hat{P}^{\delta}_{x_1} \hat{\rho}(t_1)
\hat{P}^{\delta}_{x_1} \right),
\end{eqnarray}
where we denoted $\hat{P}_x^{\delta} =
\int_{x-\delta/2}^{x+\delta/2} dy |y \rangle \langle y|.$

 To construct the probabilities
$p_{\delta}(U_1,0;U_2,t)$, we split each set $U_i$ into mutually
exclusive cells $u_{\alpha i}$ of size $\delta$, such that
\begin{eqnarray}
\cup_{\alpha} u_{\alpha i} &=& U_i \\
u_{\alpha i} \cap u_{\beta i} &=& \emptyset, \alpha \neq \beta.
\end{eqnarray}
If we denote  select points $x_{\alpha i} \in u_{\alpha i}$, for
all $i$ ($x_{\alpha i}$ may be the midpoint of $u_{\alpha i}$), we
may construct
 the probability $p_{\delta}(U_1,0;U_2,t)$
 \begin{eqnarray}
p_{\delta}(U_1,0;U_2,t) = \sum_{\alpha} \sum_{\beta}
p_{\delta}(x_{\alpha i},0 ; x_{\beta j}, t)
 \end{eqnarray}

In the limit that the typical size of the sets $U_1, U_2$ is much
larger than $\delta$, we obtain

\begin{eqnarray}
p_{\delta}(U_i,t_1|U_j, t_2) = \frac{1}{\delta^2} \int_{U_1} dx_1
\int_{U_2} dx_2 p_{\delta}(x_1,t_1;x_2, t_2), \label{additive}
\end{eqnarray}
In other words, the objects
$\frac{1}{\delta^2}p_{\delta}(U_i,t_1|U_j, t_2)$ play the role of
probability densities.

Equation (\ref{additive}) suggests that the two-time probabilities
are given by the positive operators
\begin{eqnarray}
\hat{\Pi}(U_1,0; U_2, t) = \frac{1}{\delta^2} \int_{U_1} dx_1
\int_{U_2} dx_2 \hat{P}^{\delta}_{x_1} \hat{Q}_{x_2}{\delta}
\hat{P}_{x_2}^{\delta}, \label{general}
\end{eqnarray}

which fail to define a POVM, because they are not normalized to
unity: they differ from $1$ by a term of order $O(\delta)$. This is
an artefact of the way we implemented the continuous limit in going
from  probabilities (\ref{additive}) to those of (\ref{general}).
An error of the order of $O(\delta)$ is reasonable, since the
sampling error is itself of the order of $\delta$.

It is easy to remedy this problem by working with POVM's for the
single-time probabilities. We consider a POVM $\hat{\Pi}^{\delta}
(U) = \int_U dx \hat{\Pi}^{\delta}_x$ for position that satisfies
the following properties
\begin{eqnarray}
\int dx \; \hat{\Pi}_x^{\delta} = \hat{1}, \hspace{2cm} \int dx \;
x \; \hat{\Pi}_x^{\delta} = \hat{x}.
\end{eqnarray}
For example one may consider the Gaussian POVM
\begin{eqnarray}
\hat{\Pi}_x^{\delta} = \int d\bar{x} \frac{1}{\sqrt{2 \pi} \delta}
e^{- (x- \bar{x})^2/2 \delta^2} |\bar{x} \rangle \langle \bar{x}
|.
\end{eqnarray}

Then the operators
\begin{eqnarray}
\hat{R}^{\delta}(U_1,0; U_2,t) = \int_{U_1} dx_1 \int_{U_2} dx_2
\sqrt{\hat{\Pi}_{x_1}} e^{i \hat{H}t} \hat{ \Pi}_{x_2}^{\delta}
e^{-i \hat{H}t} \sqrt{\hat{\Pi}_{x_1}} \label{R}
\end{eqnarray}
satisfy all properties of a POVM including  the normalization
condition. It is easy to check that within an error of $O(\delta)$
the probabilities defined by the POVM (\ref{R}) coincide with those
of (\ref{general}) (with $\sqrt{2 \pi} \delta$ in place of
$\delta$). The generalisation to $n$-time measurements is
straightforward
\begin{eqnarray}
\hat{R}^{\delta}(U_1,t_1; U_2, t_2; \ldots U_n,t_n) = \hspace{5cm} \nonumber \\
\int_{U_1} dx_1 \int_{U_2} dx_2 \ldots \int_{U_n} dx_n e^{i
\hat{H}t_1} \sqrt{\hat{\Pi}_{x_1}} e^{i
\hat{H}(t_2-t_1)}\sqrt{\hat{\Pi}_{x_2}} \nonumber \\ \ldots e^{i
\hat{H}(t_n-t_{n-1})} \hat{\Pi}_{x_n} e^{-i \hat{H}(t_{n}
-t_{n-1})} \ldots \sqrt{\hat{\Pi}_{x_2}} e^{- i \hat{H}(t_2-t_1)}
\sqrt{\hat{\Pi}_{x_1}} e^{-i \hat{H}t_1}. \label{Rn}
\end{eqnarray}

\subsection{Basic features of the constructed POVMs}
\subsubsection{The inequivalence with stochastic processes}
The probability densities defined by the sequence of the POVMs
(\ref{Rn}) for all values of $n$ as

\begin{eqnarray}
p_n^{\delta}(x_1,t_1; x_2, t_2; \ldots ; x_n, t_n) = Tr
[\hat{\rho}_0 \hat{R}^{\delta}(U_1,t_1; U_2, t_2; \ldots
U_n,t_n)]. \label{pn}
\end{eqnarray}

This result in conjunction with the theorems of section 3.3,
demonstrate the inequivalence of quantum probabilities for
multi-time measurements with those that can be obtained by a
classical stochastic processes. The sequence (\ref{pn}) does not
define a probability measure on the space of paths, because  the
compatibility condition necessary for the definition of such a
measure
\begin{eqnarray}
p_{n-1}^{\delta}(x_1,t_1; \ldots; x_{i-1}, t_{i-1}; \ldots ;
x_{i+1}, t_{i+1}; \ldots; x_{n}, t_{n}) = \nonumber \\
 \int dx_i
\; p_n^{\delta} (x_1,t_1; \ldots ; x_i, t_i; \ldots ; x_n,
 t_n)\label{compatibility}
\end{eqnarray}
for all possible $i = 1, 2, \ldots, n$, is not satisfied.

 A weaker version  is satisfied instead,
\begin{eqnarray}
p_{n-1}^{\delta}(x_1,t_1; x_2, t_2; \ldots ; x_{n-1}, t_{n-1}) = \hspace{3cm} \nonumber \\
 \int dx_n \; p_n^{\delta} (x_1,t_1; x_2, t_2; \ldots
; x_{n-1}, t_{n-1}; x_n, t_n), \; \; t_n > t_{n-1} > \ldots
> t_2
> t_1 \hspace{2cm} \label{Kolweak},
\end{eqnarray}
namely only if we integrate over the variables defined at the {\em
final moment} of time in the $n$-time distribution, do we obtain
the $n-1$-time probability distribution.

\subsubsection{Strong dependence on the apparatus's
resolution}

Since the functions (\ref{pn}) provide a well defined system of
joint probability densities, one could consider defining an
generalisation of stochastic processes that would reproduce the
predictions of quantum measurements. There is however a problem:
the POVM (\ref{Rn}) and the corresponding probability densities
depend very strongly on the parameter $\delta$.

This is a direct consequence of the fact that the  probabilities
(\ref{pn}) arise out of the non-additive measure (\ref{PROB}).
  Suppose we consider two different
measurement devices, one characterized by a value $\delta$ and
another by a value $2 \delta$. In any reasonable measurement scheme
one would expect that the two-time probabilities for sample sets
$U_1$ and $U_2$ would not be appreciably different if their size is
much larger than $\delta$. However, this turns out not to be the
case.  It is easier to see this in the discredited expression
(\ref{fine}).

A projection operator $\hat{P}^{2\delta}_x$ centered around $x$
with width $2 \delta$  can be written as the sum
$\hat{P}^{\delta}_{x - \frac{\delta}{2}} + \hat{P}^{\delta}_{x +
\frac{\delta}{2}}$. When we construct the elementary probabilities
corresponding to $\hat{P}^{2 \delta}_x$, which correspond to sets
of width $2 \delta$ they will differ from the probabilities for
the same sets, when the latter are constructed by sets of width
$\delta$. Their difference will be the interference term
\begin{eqnarray}
2 \, Re \, d_{\delta}(x_1 +\delta/2, x_1 - \delta/2,t_1: x_2, t_2)
=
 \nonumber \\ \, 2 Re \, Tr \left(
e^{i\hat{H}(t_2-t_1)} \hat{P}^{\delta}_{x_2}
e^{-i\hat{H}(t_2-t_1)} \hat{P}^{\delta}_{x_1+\delta/2}
\hat{\rho}(t_1) \hat{P}^{\delta}_{x_1-\delta/2} \right).
\label{interf}
\end{eqnarray}

The modulus of the 'interference' term is, in general, of the same
order of magnitude with the probabilities $p_{\delta}$ and
$p_{2\delta}$ themselves. Hence, when we sum (or integrate) over the
probabilities corresponding to the cells of width $\delta$ or $2
\delta$ to construct the probabilities
 $p_{\delta}(U_1,t_1|U_2, t_2)$ and $p_{2\delta}(U_1,t_1|U_2,
t_2)$ for generic large sample sets $U_1$ and $U_2$ the results
differ by an amount of

\begin{eqnarray}
\epsilon_{\delta}(U_1, t_1; U_2, t_2) = Re \, \int_{U_1} dx_1
\int_{U_2} dx_2 \, d_{\delta}(x_1 + \delta/2, x_1 - \delta/2, t_1:
x_2, t_2).
\end{eqnarray}
This term is of the same order as the probabilities themselves
\cite{Ana04a}, a fact pointing to the strong dependence of the
results on the resolution $\delta$.  Different values of $\delta$
lead  to very different probabilities.

We may also see that in the POVM (\ref{R}). For the special case
of a free particle $\hat{H} = \frac{\hat{p}^2}{2m}$ (\ref{R})
equals
\begin{eqnarray}
\langle x|\hat{R}^{\delta}(U_1,0; U_2,t)|x' \rangle = \frac{m}{(2
\pi)^{3/2} t \delta} \int_{U_1} dx_1 \int_{U_2} dx_2 \;
\exp \left(-i\frac{m}{t}(x-x') (x_2 - \frac{x+x'}{2}) \right) \nonumber \\
 \times \exp \left(- \frac{1}{2}(\frac{m^2 \delta^2}{ t^2} + \frac{1}{4
\delta^2})(x-x')^2 - \frac{(x_1 - \frac{x+x'}{2})^2}{2
\delta^2}\right) \hspace{2cm}\label{R2free}
\end{eqnarray}

This can be written as
\begin{eqnarray}
\langle x|\hat{R}^{\delta}(U_1,0; U_2,t)|x' \rangle = \frac{m}{ t}
\, \tilde{\chi}_{T_{\frac{x+x'}{2}} U_2} (m \frac{x-x'}{t}) \,
\chi_{U_1}^{\delta}(\frac{x+x'}{2}) \nonumber \\
\times \exp \left(- \frac{1}{2}(\frac{m^2 \delta^2}{ t^2} +
\frac{1}{ 4 \delta^2})(x-x')^2 \right), \label{Ralt}
\end{eqnarray}
where $T_x$ denotes the translation operator on ${\bf R}$,
$\tilde{\chi}$ is the Fourier transform of the characteristic
function $\chi$ and $\chi^{\delta}_U$ is a smeared characteristic
function of position. If the size of $U_1$ is much larger than
$\delta$, then one may approximate $\chi^{\delta}_U$, by an exact
characteristic function, thus obtaining
\begin{eqnarray}
\langle x|\hat{R}^{\delta}(U_1,0; U_2,t)|x' \rangle \simeq
\frac{m}{ t} \, \tilde{\chi}_{T_{\frac{x+x'}{2}}U_2}(m
\frac{x-x'}{t})\, \chi_{U_1}(\frac{x+x'}{2}) \nonumber \\
\times \exp \left(- \frac{1}{2}(\frac{m^2 \delta^2}{ t^2} +
\frac{1}{4 \delta^2})(x-x')^2 \right).
\end{eqnarray}
The matrix elements of the POVM involve a product of two terms that
depend on the sample sets (and not on $\delta$) with a Gaussian term
that is very sensitive on $\delta$ and does not depend on the sample
sets. This clearly demonstrates the strong dependence of the POVM
(\ref{R2free}) on $\delta$, which persists even for very-coarse
sample sets. It is easy to verify that the norm of the difference
between two positive operators $\hat{R}^{\delta}(U_1,0; U_2,t)$ and
$\hat{R}^{\delta'}(U_1,0; U_2,t)$, for different values  $\delta$
and $\delta'$ respectively, is of the order of the norm of the
operators themselves. For most states therefore the probabilities
will be very sensitive on the resolution $\delta$.

Note also that as $\delta \rightarrow 0$ the POVM (\ref{Rn}) does
not converge to a PVM that provides an ideal value for
probabilities, as is the case in single-time measurements. Instead,
$\lim_{\delta \rightarrow 0} \hat{R}^{\delta}(U_1,0; U_2,t) = 0$.
This behavior is a consequence of the use of the square root of the
POVM $\hat{\Pi}$ in (\ref{Rn}). This is necessary in order to
guarantee that (\ref{Rn}) is a proper POVM, normalized to unity and
with the correct dimensions to define a probability density on ${\bf
R}^n$.

\subsection{Quantum measuring device} Our previous derivation of
the probabilities for multi-time measurements was based on an
operational description of the quantum measurement process, namely
the assumption that the measuring device is classical and that the
effect of the measurement is the "reduction of the wave packet
rule", either corresponding to ideal measurements, or to non-ideal
ones (in which case we employ POVM's).

One may however object that a full quantum mechanical treatment of
the measuring device may presumably lead to a different result. We
shall argue here that this is not the case. It is well known that
the standard treatment of a quantum measurement device (together
with von Neumann's reduction rule) is equivalent to the description
of the probabilities for a quantum system with a PVM \cite{vNeu}.
For observables with continuous spectrum measurements are usually
unsharp and the sampling of the quantum system turns out to be
equivalent to a POVM that depends on specific properties of the
interaction between the quantum system and the measurement device.
This dependence is, however, relatively weak as the probability for
sufficiently coarse-grained sets is largely insensitive to such
details \cite{BLM96, BL96}.

The generalisation of the results above for sequential measurements
is straightforward. One only needs to  introduce a different
measurement device for each measurement. If the devices are
initially uncorrelated, it is easy to demonstrate that in ideal
 measurements (and discrete pointers) probabilities are provided by a POVM of the type (\ref{POVM2}).
However, if we employ a discrete pointer for the measurement of
position the dependence of probabilities on the resolution arises
out of the explicit correlation between the continuous variable
$\hat{x}$ and the discrete basis  for the pointer. Effectively we
return to a POVM like (\ref{fine}) with the same strong dependence
on $\delta$.

A continuous pointer can be shown to lead to an equation of the form
(\ref{R}), with the smearing function determined by  the initial
state of the apparatus. To see this, we consider the following  toy
model. Let $\hat{x}$ by the position of the particle we want to
determine, and let the particle be prepared in a state $|\psi_0
\rangle$. We assume two identical measurement devices each in state
$|\Psi_0 \rangle$, initially uncorrelated with the particle and with
each other. The state of the total system will then be initially
$|\psi_0 \rangle |\Psi_0 \rangle |\Psi_0 \rangle$. The pointer
variables are $\hat{q}_1$ and $\hat{q}_2$ for each device. We assume
that the self-dynamics of the devices is negligible, that the
particle's Hamiltonian is $\hat{H}$ and that  the interaction
Hamiltonian is
\begin{eqnarray}
H_{int} = f_{t_1}(t) \hat{x}\otimes \hat{k}_1 \otimes \hat{1} +
f_{t_2}(t) \hat{x} \otimes \hat{1} \otimes \hat{k}_2,
\end{eqnarray}
where $\hat{k}_i$ are the conjugate momenta of $\hat{q}_i$, and
$f_{t_i}(t)$ is a function of time sharply concentrated around
$t_i$. At the limit of instantaneous measurements the state of the
system at time $t=t_2$ is
\begin{eqnarray}
|\psi_{tot}(t_2) \rangle = \int d k_1 \int d k_2 \; \left( e^{-i k_2
\hat{x}} e^{-i \hat{H}(t_2 - t_1)} e^{-i k_1 \hat{x}} e^{-i\hat{H}
t_1} |\psi_0
\rangle \right) \nonumber \\
\otimes |k_1 \rangle \langle k_1|\Psi_0 \rangle \otimes |k_2 \rangle
\langle k_2| |\Psi_0 \rangle.
\end{eqnarray}

The probability distribution for the pointer variables is
\begin{eqnarray}
\int dx \; |\langle x, q_1, q_2 | \psi_{tot}(t_2) \rangle|^2.
\end{eqnarray}
This  coincides with that given by the POVM  (\ref{R}), if we
identify $\sqrt{\hat{\Pi}}_y = \int \frac{d k}{\sqrt{2 \pi}} e^{-i
k(\hat{x} - y)} \langle k| \Psi_0 \rangle$. It is easy to verify
that the corresponding  $\hat{\Pi}_y$ defines a POVM for a
single-time measurement of position. In this simple model the
resolution $\delta$ of the device is determined by the spread in
momentum of the apparatus's initial state. In a more realistic
model, the resolution $\delta$ receives contributions from the
self-dynamics of the detector (and of a possible environment), from
the finite duration of the measurement interaction and from the
accuracy in the readings of $\hat{q}_1, \hat{q}_2$.

 The presence of an environment does not change the essence of
the arguments presented previously. The consideration of an enlarged
Hilbert space that also contains the degrees of freedom of the
environment does not make any difference to the mathematical
formulation of the issue. The only possible way to cancel the strong
dependence of the probabilities on the resolution $\delta$, is to
assume that the effect of the environment causes interference terms
like (\ref{interf}) (where now the projectors refer to the values of
the pointer rather than those of the measured particle) to become
rapidly small\footnote{ The environment is coupled to the measuring
apparatus and not directly to the particle. If that were not the
case the particle would exhibit fully classical behavior and its
measurement would not be different from that of a classical
probabilistic system.}.

In general, the interference terms (\ref{interf}) can only be
suppressed (for a sufficiently generic initial state),  if the
environment causes the reduced density matrix of particle+apparatus
to be diagonalisable in the factorized basis $|i \rangle |a_i
\rangle$, where $|i \rangle$ are the eigenstates of the measured
observable and $|a_i \rangle$ the pointer basis in the apparatus's
Hilbert space. This is in general not possible as can be seen from a
very general theorem \cite{BG00} (and in a different but related
context by \cite{ BLM96}). Indeed if this diagonalization took place
we would have a full resolution of the so-called
macroobjectification problem by environment-induced decoherence,
which is known not to be the case\footnote{Whether
environment-induced decoherence solves the full measurement problem
in interpretations other than the Copenhagen one (e.g. many-worlds
or consistent histories) is a different issue, unrelated to the aims
of this paper. For the purposes of the present argument, we are only
interested in the mathematical statement that the diagonalization in
the basis $|i \rangle |a_i \rangle$ cannot implemented in any closed
system (however large) that evolves unitarily.}
---see the discussion in \cite{Schl04, Adl01, Das05}.

Hence we conclude that the strong dependence of the multi-time
probabilities on the properties of the measurement device is
unavoidable, whether one considers the formalism of quantum theory
as an operational description, or whether one considers any
minimal generalisations that involve a quantum mechanical
treatment of the measuring device.

\subsection{Consequences}
\subsubsection{Contextuality of measurements}
The first result of our analysis of sequential measurements is the
breakdown of quantum logic. Unlike the single-time case, a
proposition about a measurement outcome is not represented by a
projection operator. The probabilities $p(U_1,t_1;U_2,t_2)$ do not
depend on the sample sets through projectors and  are very
different in different experimental setups. A two-time YES-NO
experiment will lead to different probabilities from  those
obtained by the experiment of Fig. 3.

This result is complementary to the Kochen-Specker theorem: it is
in general not possible to attribute specific values to sets of
observables, even commuting ones, without specifying the context,
namely the concrete experimental set-up. In  other words, one
cannot define a sample space for the possible outcomes of an
observable, without referring to the specific measurement being
implemented.

The relevance of contextuality in sequential measurements (both
factual and counterfactual) has been studied extensively in the
literature. Albert, Aharonov and D' Amato  employed the result of
Ref. \cite{ABL64} concerning an ensemble that is both pre- and
post-selected through measurements at times $t_i$ and $t_f$
\cite{AAD85a}. They showed that it is possible to retrodict the
results of specific measurements that could have been  carried out
at any moment of time in the time interval $[t_i, t_f]$. It is
important to remark that this retrodiction can be applied to
incompatible measurements, i.e. ones corresponding to non-commuting
observables. A similar result is also obtained by Kent \cite{Kent},
who  argues that retrodiction in a quantum theory that purports to
describe individual systems (consistent histories)  leads
generically to contrary inferences.

The discussion in this paper is within a slightly different context
than the ones of the references above. We are only interested in
providing a probability measure for the results of sequential
measurements that have actually taken place. The intermediate
measurement device is part of a specific experimental set-up and we
do not consider any counterfactual statements (about retrodiction).
In any case, the situations studied here typically involve
probabilities that are spread over different alternatives: typically
only trivial inferences can be made.

The standard proof of the Kochen-Specker theorem assumes that it is
possible to assign definite values to commuting physical observables
in individual systems prior to measurement, which, while reasonable,
is not an statement amenable to empirical verification. Moreover, it
involves an interpretation-dependent assumption that it is possible
to extrapolate the rules of quantum theory from the description of
statistical ensembles to that of individual systems. In sequential
measurements however the dependence of the measurement outcomes on
the specific experiment is direct in terms of concrete empirical
data, and makes no assumptions other than that quantum theory
provides the correct probabilities for the measurement outcomes in
statistical ensembles.

Contextuality can also be inferred from Bell's theorem and its
generalisations or from Wigner's theorem about the lack of a joint
probability distribution for non-commuting observables.
 However, in the cases above one may provide alternative
explanations: in the former case one may attribute the failure of
Bell's inequalities to non-locality, while in the second one may
invoke the inability to perform simultaneously measurements of
incompatible observables. There are no such limitations,  when the
argument for quantum contextuality is phrased in terms of the
probabilities for sequential measurements. It is in principle
possible to measure multi-time probabilities in different
experimental set-ups. If the results of our analysis are correct,
these probabilities will differ strongly, thus providing irrefutable
empirical evidence about the contextuality of quantum events.

\subsubsection{Inferences and conditional probability}
Conditioning it is a very important part of classical probability;
 it is the mathematical implementation of the idea that when we obtain
 information from an experiment, we need to modify our description of  the system
 (i.e. the probability distribution) in order to account for the new information.
 The prototype of conditioning  is the notion of conditional
 probability, i.e. the probability $p(A|B)$ that $A$ will take place when we have
 verified that $B$ occurred
\begin{equation}
p(A|B) = \frac{p(A \cap B)}{p(B)} \label{condprop}
\end{equation}

It is sometimes suggested that the "wave packet reduction rule" can
be interpreted as a quantum version of conditional probability. Our
results suggest that this is not the case. Such an interpretation is
only possible, when 'conditioning' refers to the most fine-grained
recordings of a physical system's properties. If we attempt to
employ  this rule to account for coarser alternatives, we inevitably
lose information in the process and cannot obtain correct physical
predictions.

In the classical theory conditional probability can be employed to
define logical implication. If the conditional probability $p(A|B)$
for an event $A$ given that  $B$ was realized equals $1$. If in an
experiment we verify the property $B$, we may expect that the
property $A$ will be {\em almost} surely satisfied. In quantum
theory the situation concerning implication is more subtle. There is
a strong distinction between prediction and retrodiction.

The fact that Eq. (\ref{compatibility}) does not hold implies that
retrodiction is problematic. A two-time measurement involves a
different experimental set-up from that of a single-time
measurement. The single-time probability for a measurement at time
$t$ is different from any marginal  obtained by tracing out the
results of any measurement at any time $t' < t$. It is, therefore,
impossible to make any inference (or any probabilistic statement)
about what would have happened if a measurement had taken place
earlier, {\em solely} from the data obtained at time $t$. It is
necessary to provide the specifics of the intermediate measurement
scheme. On the other hand, the validity of Eq. (\ref{Kolweak})
implies that prediction works the same as in the classical case.
Tracing out the results of a later measurement yields the same
probability distribution as if the measurement had not taken place,
hence it is in principle possible to make inferences from the data
at time $t$ about the results   of  measurements that takes place at
time $t'
> t$.

\subsubsection{Strong dependence on the measuring device's resolution}
In single-time quantum theory we know that different experiments
of the same type are expected to yield identical results, up to
sampling and systematic errors. This is guaranteed by the spectral
theorem, and it is epistemologically very desirable because  the
results of similar experiments can be immediately compared. But in
multi-time measurements even two experiments that are identical in
all details (preparation of the measuring device, source of
particles, design of the experiment) but the resolution of the
measuring device, will lead to different probabilities and
correlation functions. This is a very stronger effect and in
principle observable. It is a source of doubt about whether any
meaningful information can be obtained from such experiments.

At a practical level we know that experiments yield more reliable
results, when we have expended time and effort to minimis the
errors, which may arise from either sampling inaccuracies or from
the finite resolution of the measurement device. Copenhagen quantum
theory assumes that any results we obtain will make reference to the
specific set-up, but even when we restrict our expectations to that
case, common sense suggests that the smaller the error, the more
reliable our experimental results will be. The event frequencies
should converge to some ideal values that would characterize, if not
the measured system in itself, at least the general design of the
experiment. This expectation is fulfilled in single-time quantum
theory. The dependence of the typical POVMs for unsharp measurements
on the error (or resolution) $\delta$ is rather weak, and for
sufficiently coarse-grained samplings the corresponding positive
operators are close to true projectors. The probabilities
corresponding to  such samplings probabilities will  therefore be
the same in all measurement devices, and they will coincide with the
probabilities obtained from YES-NO experiments. But in multi-time
measurements this is no longer  the case. The POVM's dependence on
$\delta$ persists even for very coarse samplings. When we increase
the resolution, we do not obtain "better" results, we simply obtain
different results.

\section{Sequential measurement in hidden-variable theories}

\subsection{Bohmian mechanics}
 The results of section 3.5 suggest that the probabilities for sequential measurements are very sensitive to
 even minute changes  of the measuring procedure. Hence even
if we accept that such probabilities can be defined from the
statistical data, a question is immediately raised. Why are
multi-time quantum measurements so different from single-time
ones? Clearly, an answer to this question cannot be provided
within standard quantum theory, because the issue itself arises as
a consequence of the theory's basic postulates. One would have to
enlarge the domain of standard quantum theory and essentially work
with hidden variables.

The most important hidden variables theory, both because of its
long history and its intrinsic strength, is Bohmian mechanics \cite{BH}. In this
theory the additional variables are the particles's position,
which evolve according to the modified Newton's equations
\begin{eqnarray}
m \dot{x} = Im \frac{\partial_x \Psi(x,
t)}{\Psi(x,t)}\label{Bohmeom},
\end{eqnarray}
where the wave function $\Psi(x,t)$ is a solution of Schr\"odinger's
equation.

The dynamical equations (\ref{Bohmeom}) are usually supplemented by
the condition of {\em quantum equilibrium}, namely that in a
statistical ensemble of particles the probability density for the
particle's position  is given by Born's rule: $\rho(x,t) = |\Psi(x
,t)|^2$.

Assuming quantum equilibrium, it is easy to construct a
probability density for the outcomes of sequential measurements.
The particle's position and the wave function satisfy a set of
differential equations and are fully deterministic. Hence any
trajectory can be fully specified by the knowledge of the initial
conditions: the position $x_0$ and the wave function $\Psi_0(x)$.
Let us denote as $x(t; x_0, \Psi_0]$ the solutions to
(\ref{Bohmeom}); they can be viewed as functions of the random
variable $x_0$.

The probability that the particle lies in the set $U_1$ at time
$t_1$, in $U_2$ at time $t_2$... and in $U_n$ at time $t_n$ then
equals
\begin{eqnarray}
p^n(U_1,t_1;U_2,t_2; \ldots ; U_n, t_n)) = \int dx_0
|\Psi_0(x_0)|^2 \hspace{3cm}\nonumber \\
\times \chi_{U_1}\left[x(t_1;x_0,\Psi_0] \right]
\chi_{U_2}\left[x(t_2;x_0,\Psi_0] \right] \ldots
\chi_{U_2}\left[x(t_n;x_0,\Psi_0] \right]. \label{BohmPn}
\end{eqnarray}
 The tower of
all $n$-time probabilities satisfy by construction the compatibility
condition and thus defines a measure on the space of paths $d
\mu[x(\cdot)]$, which depends only on the initial wave-function and
the Hamiltonian operator of the wave function's evolution. This
measure fully reproduces the predictions of standard quantum theory
at a single moment of time.

There is no contradiction with our results at section 3.3. The
non-go theorems proved there refer to probabilities that can be
constructed via POVMs: the n-time probability densities are  {\em
linear} functionals of the initial density matrix. Clearly, this is
not the case here as the initial state enters in a non-trivial way
in the definition of the random variables $x(t;x_0,\Psi_0]$. For the
same reason the stochastic process corresponding to (\ref{BohmPn})
is non-Markovian\footnote{If we distinguish the two roles of the
wave function as probability distribution and as agent of dynamical
evolution, then the stochastic process is Markovian. It is however
not time-homogeneous, unless the wave function is an eigenstate of
the Hamiltonian operator.}. It is however local-in-time, because the
densities corresponding to (\ref{BohmPn}) factories
\begin{eqnarray}
p^n(x_1,t_1; x_2, t_2; \ldots; x_n,t_n) = |\Psi(x,t_1)|^2
\hspace{4cm} \nonumber \\ \times  \delta(x_1,
g_{t_1,t_2}[\Psi_0](x_2)) \delta(x_2, g_{t_2,t_3}[\Psi_0](x_3))
\ldots \delta(x_{n-1}, g_{t_{n-1},t_n}[\Psi_0](x_n)),
\end{eqnarray}
where $g_{t,t'}[\Psi_0], t < t'$ is the backwards-in-time
evolution operator corresponding to the equation of motion
(\ref{Bohmeom}).

It would seem from the above expression that  multi-time
probabilities in Bohmian mechanics are different from those of
standard quantum theory. The multi-time probability distributions in
standard quantum theory cannot be obtained from a probability
distribution. A difference in the probabilistic outcomes of
multi-time measurements between Bohmian mechanics and quantum theory
has been suggested before in \cite{Neu00, FPSW02}. In these
references the predictions of Bohmian mechanics were compared with
correlation functions of the form (\ref{cf}), which have no
immediate operational interpretation, while here we compare them
with the probabilities for sequential measurements, which can in
principle be determined empirically. Our analysis also shares some
features with that of Hartle \cite{Har04}.

An immediate objection can be raised to the analysis above.
 Bohmian mechanics  refers to the properties of things in themselves and not to
measurement outcomes. To obtain the measured probabilities one would
have to model the interaction of the quantum system with a measuring
device. The Bohmian description of quantum measurements has been
developed in \cite{Bohmmeas}--see also a related discussion about
Stochastic Mechanics in \cite{Stochmeas}. In these references it is
argued that the reduction of the wave packet rule can be obtained
from Bohmian mechanics after the interaction of a system with a
measuring device has been taken into account. One would therefore
expect that in  sequential measurements the predictions of quantum
theory should be reproduced.

We next examine this issue in more detail. We consider a two time
measurement. Let $x$ be the particle's position, $Q^1$ and $Q^2$ the
variables for the first and second measurement device respectively,
and $X$ a pointer function of $Q_1$ or $Q_2$, the range of which is
the space $\Omega$ of possible alternatives in each measurement. The
pointer function may be either continuous or discrete.  The total
system will be effectively described by a stochastic process
analogous to the one defined by (\ref{BohmPn}). Assuming quantum
equilibrium for the total system,  the probabilities that the
pointer $X$ is found in a set $U_1$ at $t_1$, and in a set $U_2$ at
$t_2$ equals

\begin{eqnarray}
p^2(U_1,t_1;U_2,t_2) = \int \,dx_0 \, dQ^1_0 \, dQ^2_0 \,
|\Psi_0(x_0, Q^1_0, Q^2_0)|^2 \hspace{3cm}\nonumber \\
\times \chi_{U_1}\left[X(Q^1(t_1;x_0, Q^1_0, Q^2_0\Psi_0]) \right]
\chi_{U_2}\left[X(Q_2(t_2;x_0,  Q^1_0, Q^2_0, \Psi_0]) \right]
\label{Bohmmeas}
\end{eqnarray}
where $Q^i(t, x_0, Q_0, \Psi)$ are the solutions to the
deterministic equations of motion for the variables $Q$ written in
terms of the initial condition. The marginal probability, in which
the results of the first measurement have been traced out equals
\begin{eqnarray}
p^2(\Omega, t_1; U_2, t_2) = \int \,dx_0 \, dQ^1_0 \, dQ^2_0 \,
|\Psi_0(x_0, Q^1_0, Q^2_0)|^2 \nonumber \\
\times \chi_{U_2}\left[X(Q_2(t_2;x_0, Q^1_0, Q^2_0, \Psi_0])
\right]. \label{B1}
\end{eqnarray}
On the other hand the probability of a single-time measurement at
time $t_2$ is
\begin{eqnarray}
p^1(U_2, t_2) = \int \,dx_0 \, dQ^2_0 |\Psi_0(x_0, Q^2_0)|^2
\chi_{U_2}\left[X(Q_2(t_2;x_0, Q^2_0, \Psi_0]) \right]. \label{B2}
\end{eqnarray}
The crucial difference lies  in the equations of motion for the
pointer variable--there are no $Q_1$ variables in the expression for
$p^1$, because there is no measuring device at time $t_1$. The
probabilities (\ref{B1}) and (\ref{B2}) refer to different physical
systems and as such they correspond to a different stochastic
process.

\subsection{Contextuality from non-locality}

The analysis of multi-time probabilities in Bohmian mechanics above
does not guarantee that the predictions of Bohmian mechanics
coincide with those of standard quantum theory. We shall demonstrate
now that the key property that permits that is the inherent
non-locality of Bohm's theory. For  a different derivation of the
constraints from local realism to the conditional probabilities of
sequential measurements the reader is referred to \cite{Pop94,
P3H98}. Also related are the constraints that can be expressed in
terms of "temporal Bell inequalities" \cite{ LeGa86,PaMa93, CCO99}.
The derivation we provide here refers to the most general case.

We consider a general deterministic hidden variable theory. The
probabilistic description arises from an initial probability
distribution for a statistical ensemble, which is related to the
wave functions by Born's rule. To model a sequential measurement
 we assume the same variables $x$, $Q^1$ and $Q^2$ as in section
 4.2. The wave function $\Psi$ at $t=0$ is assumed factorized, namely
 $\Psi_0(x_0, Q^1_0, Q^2_0) = \psi_0(x_0)
\phi_1(Q^1_0) \phi_2(Q^2_0)$. We also assume that the degrees of
freedom $Q_1$ and $Q_2$ do not interact directly, and that at time
$t_1$ the particle has not interacted with the degrees of freedom of
the second measurement device. This implies that the value of $x$
and $Q^1$ at time $t_1$ does not depend on the initial value
$Q^2_0$. The conditions above are natural in any measurement
process.

We  first consider a discrete pointer $X$, that takes values in the
finite set $\Omega$ of elementary alternatives. We assume that after
a measurement the pointer $X$ reveals the value of the function $f$
of the variable $x$, so that at the time $t$ when the measurement
interaction has finished
\begin{eqnarray}
X[Q(t, x_0, Q_0] = f[x(t,x_0,Q_0)],  \label{cond1}
\end{eqnarray}

where $x_0, Q_0$ are the initial values of the configuration
variables of the system and apparatus respectively. We simplify
the notation by writing $X[Q(t, x_0, Q_0] = X_t(x_0, Q_0)$ and
$f[x((t,x_0,Q_0] = f_t(x_0, Q_0)$. The variables $x$ need not only
refer to particle positions, but may in principle refer to other
degrees of freedom, e.g. spin.

We next assume that the single-time predictions of this theory
coincide with those of standard quantum mechanics for ideal
measurements, namely ones corresponding to a PVM $\hat{F}_U$ on
$\Omega$ defined on the Hilbert space of the system's wave
functions.
\begin{eqnarray}
 p(U, t) = \int dx_0 \, dQ_0 \, |\psi_0|^2(x_0) \, |\phi|^2(Q_0)
 \,
 \chi_U[X_t(x_0,Q_0)] = \nonumber \\
 \int dx_0 |\psi_0|^2(x_0)
 \chi_U[f^{sys}_t(x_0)] = \langle \psi_0|e^{i \hat{H_0}t}\hat{F}_U e^{-i \hat{H_0}t}|\psi) \rangle, \label{cond2}
\end{eqnarray}
where $f_t^{sys}(x_0)$ refers to the evolution of the measured
system in absence of the measurement device (equivalent to the
quantum mechanical evolution with a Hamiltonian $\hat{H}$). Equation
(\ref{cond2}) holds in Bohmian mechanics when $f$ is a function of
position, but may be valid for more general configurational
variables. The key assumption is that the operational predictions of
quantum theory are valid, namely that the probabilities for
measurements can be obtained by an application of Born's rule in the
wave function of the system alone. This holds for ideal
measurements.

The crucial assumption is that  after the first measurement has been
completed the system does not interact any more with the first
device. The variables $Q^1$ do not appear any more in its equation
of motion. This is essentially an assumption of {\em locality} for
the interaction of the system with the measurement device. It
implies that
\begin{eqnarray}
x_{t_2}( x_0, Q^1_0, Q^2_0) = x_{t_2}(x_{t_1}(x_0, Q^1_0), Q^2_0) \nonumber \\
Q^2_{t_2}( x_0, Q^1_0, Q^2_0) = Q^2_{t_2}(x_{t_1}(x_0, Q^1_0),
Q^2_0), \label{locality}
\end{eqnarray}
namely the values of $x$ and $Q_2$ after the second measurement
depend on $Q_1$ only through the value of $x$ immediately after
the first measurement. Using the locality condition the
probability (\ref{Bohmmeas}) is written as
\begin{eqnarray}
&&p^2(U_1,t_1;U_2,t_2) = \int \,dx_0 \, dQ^1_0 \, dQ^2_0 \,
|\psi_0|^2(x_0) |\phi_1|^2(Q^1_0) |\phi_2|^2(Q^2_0) \hspace{3cm}\nonumber \\
&&\times \chi_{U_1}\left[X_{t_1}(x_0, Q^1_0)) \right]
\chi_{U_2}\left[X_{t_2}(x_{t_1}(x_0, Q^1_0), Q^2_0) \right] =
\nonumber \\
&&\int \,dx_0 \, dQ^1_0  \, |\psi_0|^2(x_0) |\phi_1|^2(Q^1_0) \;
\chi_{U_1}\left[X_{t_1}(x_0, Q^1_0) \right]
\chi_{U_2}\left[x^{sys}_{t_2}( f_{t_1}(x_0, Q^1_0)) \right] =
\nonumber \\
&&\int \,dx_0 \, dQ^1_0  \, |\psi_0|^2(x_0) |\phi_1|^2(Q^1_0) \;
\chi_{U_1}\left[X_{t_1}(x_0, Q^1_0) \right]
\chi_{(x^{sys})^{-1}U_2}\left[X_{t_1}(x_0, Q^1_0) \right] = \nonumber \\
&&\int \,dx_0 \, dQ^1_0  \, |\psi_0|^2(x_0) |\phi_1|^2(Q^1_0) \;
\chi_{U_1 \cup (x^{sys})^{-1}U_2}\left[X_{t_1}(x_0, Q^1_0))\right]
= \nonumber \\
&&\int \,dx_0 \, dQ^1_0  \, |\psi_0|^2(x_0) |\phi_1|^2(Q^1_0) \;
\chi_{U_1 \cup (x^{sys})^{-1}U_2}\left[f_{t_1}(x_0, Q^1_0)
\right]  = \nonumber \\
&&\int \,dx_0 |\psi_0|^2(x_0) \chi_{U_1 \cup
(x^{sys})^{-1}U_2}\left[f^{sys}(t_1,x_0)) \right] = \nonumber \\
&&\int dx_0 |\psi_0|^2(x_0) \chi_{U_1}\left[f^{sys}_{t_1}(x_0))
\right]
\chi_{(x^{sys})^{-1}U_2}\left[f^{sys}_{t_1}(x_0) \right]= \nonumber \\
&&\int dx_0 |\psi_0|^2(x_0) \chi_{U_1}\left[f^{sys}_{t_1}(x_0)
\right] \chi_{U_2}\left[f^{sys}_{t_2}(x_0) \right].
\label{derivation}
\end{eqnarray}
In the above derivation we employed Eq. (\ref{cond2}) in going from
the second to the third line;  in going from the third to the fourth
line we used Eq. (\ref{cond1}) and denoted by $(x^{sys})^{-1}$ the
inverse of the deterministic law of motion that takes $x_{t_1}$ to
$x_{t_2}$  in absence of apparatus; in going from the fourth to the
fifth line we used the fact that $\chi_{U_1} \chi_{U_2} = \chi_{U_1
\cap U_2}$ if $U_1 \cup U_2 = \emptyset$. In going  from the fifth
to the sixth line we used Eq. (\ref{cond1}). Note that
$f^{sys}_t(x_0)$ stands for $f(x^{sys}_t(x_0))$.

We proved therefore that the two-time probabilities coincide with
those  obtained for a stochastic process constructed solely from
the degrees of freedom of the measured system. Hence  a hidden
variables theory that satisfies (\ref{locality}) and reproduces
the single-time predictions of standard quantum theory exhibits no
contextuality in sequential measurements. Obviously $p^2$
satisfies the compatibility condition (\ref{compatibility}) and
thus differs from the corresponding predictions of standard
quantum theory.

The result above does not apply to Bohmian mechanics, because the
latter does not satisfy  the locality condition (\ref{locality}). As
the wave function evolves, it becomes entangled in the variables $x$
and $Q_1$ after the first measurement--see the discussions in
\cite{surreal}. As a result of Eq. (\ref{Bohmeom}), the equation of
motion for $x$ after the first measurement will explicitly involve
$Q_1$ and hence the Eq. (\ref{locality}) will not be satisfied. Due
to entanglement the measured system continues to be affected by the
degrees of freedom of the first measurement device even if it is far
away from it. We see therefore that what appears as strong
contextuality of the empirical probabilities in standard quantum
theory, in Bohmian mechanics it arises as a consequence of the role
of the wave function as a carrier of non-locality  in   entangled
systems.

In the Appendix B  we provide a generalisation of this result for
unsharp measurements, and also for hidden variable theories that can
be modeled by a Markov process. It is important to remark that in
deriving these results we need make no assumptions about the
explicit form of the POVM for sequential measurements: any POVM that
satisfies the assumption of Proposition 2 in Section 3.3.2 is
adequate for this purpose.

\subsection{Other hidden variable models}
Is it possible to write hidden variable theories that reproduce the
predictions of quantum theory for sequential measurements, without
violating some form of the locality assumption? This is the same
question that may be asked about hidden variable theories that
violate the Bell inequalities. If such theories are to account for
the single-time predictions of quantum theory one needs to introduce
a probability density for these variables (either fundamental or
emergent) and the usual calculus of probabilities almost guarantees
that sequential measurements will be described by a stochastic
process.

The only conceivable hidden variable theories compatible with
standard quantum theory in sequential measurements would be ones
that introduce additional variables, other than the ones necessary
to obtain agreement with the predictions of quantum theory.
 They would correspond to degrees of freedom fundamentally different from those
of classical mechanics  (e.g. 't Hooft's deterministic quantum
theory \cite{hooft}). One may then assume that these variables are
highly uncontrollable (or exhibit a kind of "coherence" within the
elements of an ensemble) so that their statistical effect cannot be
modeled by any probability density. Hence it would not be possible
to write a stochastic process for the multi-time probabilities of
the theory, and consequently the arguments of section 4.2 would not
follow. At the moment this is just a conjecture, for no such model
has been explicitly constructed. Indeed, how could one effect the
statistical descriptions of systems that are not described by
probabilities?
 However, the
existence of this possibility suggests that one might  avoid both
contextuality and non-locality by relaxing the condition that the
full system is described by a probability theory that satisfies the
Kolmogorov axioms. We shall consider this issue in the next section.

\section{Beyond probabilities}

\subsection{Motivation}

In single-time measurements probabilities are defined naturally in
terms of the projective geometry of the Hilbert space, through the
spectral theorem. This is not the case for sequential quantum
measurements. A choice of basis is necessary to take into account
the effect of the intermediate measurements. This results in a
probability assignment that is not natural with respect to  the Hilbert
space geometry (i.e. it does not preserve quantum logic).
Contextuality of probabilities follows. If one attempts to explain
it away by hidden variable models, one needs to introduce
non-local features in the interaction of the measured system with
the apparatus.

In standard quantum theory contextuality is generic, as witnessed
by the Kochen-Specker theorem. In sequential measurements however
the dependence on probabilities on even minor details of the
measurement scheme appears as rather too extreme. It is a natural
question then whether one can introduce an interpretative scheme
that avoids it, without assuming on the same time non-locality at
the fundamental level. The only way to do this, would be if the
multi-time probabilities respected quantum logic, or in other
words if they could be expressed in a way that respects the
projective geometry of the Hilbert space.

The consistent histories approach preserves quantum logic not at the
level of measurement outcomes but at the level of individual quantum
systems. The non-additivity of the measure (\ref{PROB}) is
sidestepped by assuming that probabilities can only be defined for
specific sets of histories (consistent sets), in which (\ref{PROB})
is reduced to an additive probability measure. Still, consistent
histories does not avoid contextuality, whenever one attempts to
make logical inferences based on the probabilities obtained by
different consistent sets. This is natural from a mathematical point
of view, because any probability assignment depends not only on the
projectors representing the relevant properties of the system, but
also on the consistent set.

The only conceivable way to obtain uncontextual predictions for
quantum theory would be if the non-additive measure (\ref{PROB})
could be employed for {\em  any} sampling of measurement outcomes
described by the corresponding projectors. Indeed, a key assumption
the derivation of Bell's and Kochen-Specker theorems is that a
probability distribution satisfying the Kolmogorov axioms can be
employed to model probabilities in a statistical ensemble. Hence a
quantum theory based on a non-additive measure for multi-time
measurements may potentially avoid the consequences of those
theorem--see the discussion in references \cite{Ana01a, Ana03a}.
However, empirical probabilities (that refer to the same measurement
set-up) are always additive as they correspond to relative
frequencies. The only way to relax the Kolmogorov probability
conditions is to assume that frequencies do not always define
probabilities, i.e. that they generically do not converge. The
non-additivity of the probability measure (\ref{PROB}) would then
provide an estimate of this lack of convergence.

We shall explore this rather unconventional  alternative in this
section. It may seem rather contrary to the standard use of
probabilities in quantum theory, but we believe it is worth
studying, because it is the only conceivable alternative to the
strong contextuality of standard quantum theory and non-locality.
In any case its predictions are in principle distinguishable from
those of standard quantum theory.

Before proceeding in the further examination of this hypothesis of
non-converging frequencies, we shall first provide a different
motivation for its introduction. We shall argue that it may be a
natural description of the statistical data of sequential
measurements, solely from an operational point of view.

\subsection{Sequential measurements: stability of the sample space}

While classical probability has been remarkably successful in
modeling various physical systems, its applicability to a specific
situation is not {\em a priori} guaranteed. One needs to provide
{\em physical} arguments why probability theory can model the
outcomes of a specific experiment. These arguments involve an
explanation of the choice of the variables that define the sample
space and a justification why the different runs of the experiment
define a proper statistical ensemble.

The relation of probabilities to event frequencies suggests that the
experiment can be repeated a large number of times with the same
preparation, or at least in such a way that variations in the
preparation procedure affect little the results of the experiment.
If this condition cannot be satisfied, then we cannot talk about a
statistical ensemble and have no reason to expect that the measured
frequencies would in any way allow us to determine meaningful
empirical probabilities. Small variations in the preparation and
execution of the experiment are not so much a problem: there is
always a sampling error and intrinsic uncertainty in the
determination of any experiment and we shall see in the next section
how this can be taken into account by the consideration of unsharp
measurements. If, however, the uncertainties in an experiment are
too large, there is little hope of extracting meaningful
probabilistic information from it. In other words, the use of
probability theory in modeling a physical system requires a
condition of {\em stability of the sample space}, i.e. the
assumption that the relation of the mathematical space of physical
alternatives to the experimental outcomes remains the same in all
elements of the statistical ensemble. In classical physics at least,
the above condition is a necessary requirement for any meaningful
experiment--if it is not satisfied then one usually asserts that the
corresponding experiment is ill-designed.

A key feature of the probabilities for sequential measurements we
derived in section 3 is the strong dependence on the parameter
$\delta$ that quantifies the fuzziness of single-time measurements.
In classical probability the fuzziness includes contributions of
very different physical origins: sampling and systematic error,
specific features of the measurement device and the effect of
uncontrollable parameters, whose effect cannot be reproduced
identically in all measurements of the statistical ensemble. It is
usually unnecessary to distinguish between the different
contributions: $\delta$ may be taken as an {\em upper limit} of all
possible sources of error, as it does not affect the probabilities
of sufficiently coarse samplings. In sequential measurements this is
no longer the case. If $\delta$ is considered as a measure of the
uncontrollable parameters in the system, the sensitivity of
probabilities on its value, implies that the probability density
relevant to each different run of the experiment will be
substantially different from each other \footnote{If $\delta$ is an
upper limit for the effect of uncontrollable parameters, different
sub-ensembles may be characterized by different values of $\delta$
and hence different probability assignments.}. It is difficult to
see, how a statistical ensemble of reproducible experiments is
meaningful, if their outcomes depend so strongly on the values of
uncontrollable parameters. This suggests strongly that the sample
space for sequential measurements is not stable. One therefore may
question whether  empirical probabilities can be constructed in that
case. It is quite likely, on operational grounds alone, that the
frequencies obtained do not exhibit the needed convergence
properties to define genuine probabilities.

\subsection{Non-convergent frequencies}
\subsubsection{The case of classical probability}
While the relation of probabilities to event frequencies is the
basic principle in any statistical manipulation of data, its
application is not straightforward. All statistical samples are
obtained from a finite number of experimental runs, while the
probabilities are defined from event frequencies in the limit that
the number of runs goes to infinity.
 This has been traditionally a
very strong argument against the {\em definition} of probabilities
through frequencies. However, for  practical purposes it suffices
that we consider a sufficiently large ensemble so that the
frequencies seem to stabilize. The central limit theorem guarantees
that {\em if the description of a system by classical probability
theory is valid}, then the error in the determination of
probabilities after $n$ runs will fall like $n^{-1/2}$ for
sufficiently large $n$.

More relevant to the present discussion is the behavior of relative
frequencies in unsharp measurements, namely when there is an error
of $\delta$ (sampling error or effect of uncontrollable parameters)
in the specification of probabilities. In that case the sequence
$\nu_n(U)$ of event frequencies cannot be expected to stabilize to a
probability even after a large number of runs, if the
coarse-graining scale $L$ of $U$ is of the order of magnitude of
$\delta$: sampling is simply unreliable at this scale. There will be
a region of convergence: no matter how many experimental runs we
take into consideration the sequence will take values in a region of
finite size.

We may define a quantitative measure for the failure of a sequence
to converge to a specific value. If a sequence $\nu_n$ does not
converge then  for  $n, m > N$, where $N$ may be a large integer,
we cannot find a number $\epsilon$, such that $\nu_n - \nu_m| <
\epsilon$. This suggests defining the degree of non-convergence of
$\nu_n$ as the limit
\begin{eqnarray}
\epsilon[\nu_n] = \lim_{N \rightarrow \infty} \sup_{n,m >N} |\nu_n
- \nu_m|.
\end{eqnarray}
If $\nu_n$ is a sequence of relative frequencies it is easy to
verify that $\epsilon[\nu_n] \leq 1$ and that for a converging
sequence $\epsilon[\nu_n] = 0$. Since in practice a sequence never
converges, we need to establish a rather more heuristic criterion:
we say that $\nu_n$ converges to a probability $p$, if the
parameter $\epsilon$ is much smaller than the value of $p$. In
that case it defines the size of error (or ambiguity) in the
determination of $p$.

We next examine how the ratio of convergence for a sequence of
events is related to the fuzziness of a measurement scheme.
 Let us denote by $\bar{\nu}_n$ sequence of relative
frequencies constructed from the experimental data (hence being
inaccurate due to sampling errors), and by $\nu_n$ the ideal
frequency that converges to some probability $p$, we see that
\begin{eqnarray}
| \bar{\nu}_n - p| \leq |\bar{\nu}_n - \nu_n|  + |\nu_n - p|.
\end{eqnarray}
The second term falls with $n^{-1/2}$ for large $n$, since it is
assumed to converge ideally. The first term in the right-hand-side
converges for large $n$ to
\begin{eqnarray}
\int dx \rho(x) |\chi_U^{\delta}(x) - \chi_U(x)|,
\end{eqnarray}
for some smeared characteristic function for $U$ that takes into
account the effect of sampling errors. We essentially assume that
$\bar{\nu}_n = \frac{1}{n} \sum_{i=1}^n \chi_U^{\delta}(x_n)$,
hence the ambiguity in the sampling of $x_n$ is transferred in the
smeared characteristic function. Hence, we have
\begin{eqnarray}
| \bar{\nu}_n - p| < c \frac{\delta}{L} \, \; \; n \rightarrow
\infty,
\end{eqnarray}
or for probability distributions with spread much larger than
$\delta$ we have the more stringent estimation (see Appendix A)
\begin{eqnarray}
| \bar{\nu}_n - p| < c \frac{\delta}{L} p
\end{eqnarray}


\subsection{Quantitative estimation}

In this section we explore the theoretical possibility that the lack
of
 a non-contextual probability measure for multi-time histories is
indicative of a failure of the event frequencies to stabilize after
a large number of runs. To elaborate on this proposal we  first need
to guarantee that this assumption is compatible with the single-time
predictions of quantum theory. This follows trivially from the fact
that (\ref{PROB}) is additive for single-time measurements. The same
holds for multi-time measurements, for which the projectors
$\hat{P}_{U_1}$ commute with the Hamiltonian.

We would expect the lack of convergence appear in all  $n$-time
measurements, for which (\ref{PROB}) is genuinely non-additive. The
lack of additivity is quantified by the object (\ref{decfun}),
namely the decoherence functional in the consistent histories
approach. The decoherence functional should be a measure of the
degree of non-convergence for histories. There is another argument
that lends plausibility to this expectation. The absolute values of
the off-diagonal elements of the decoherence functional often become
very small, when the selected histories are sufficiently
coarse-grained. It then becomes a good approximation to assign
probabilities to such histories. This situation can be compared with
the behavior of relative frequencies under coarse-graining. If our
sampling is of the order of the measurement error $\delta$,
frequencies do not stabilize to probabilities. If we coarse-grain
sufficiently however, so that the sample set is of size $L>>\delta$,
the relative error falls like $\frac{\delta}{L}$ and reasonable
probabilities can be approximately defined.

The analogy above is only mathematical, as the postulated lack of
convergence in quantum theory cannot be explained away as
measurement error. It strongly argues however that the decoherence
functional should encode the information of the frequency's
non-convergence. In effect, one may assign a probability for a
specific sampling $U_1, U_2, \ldots, U_n$, if the decoherence
functional between the corresponding history $\alpha$ and its
negation $\not \alpha$ (corresponding to the subset $\Omega^n -
U_1 \times U_2 \times \ldots \times U_n$ ) is much smaller in
magnitude than the probability associated to $U_1, U_2, \ldots,
U_n$, or in other words if
\begin{eqnarray}
2 \,Re \, d(\alpha, \not \alpha)  << d(\alpha, \alpha),
\label{condition}
\end{eqnarray}

This suggests that the proper measure for the relative rate of
convergence $\epsilon[\nu_n]/p$ should be identified with the
ratio $\frac{Re d(\alpha, \not \alpha )}{d(\alpha, \alpha)}$.

We need to comment at this point on the difference of the present
proposal from the consistent histories approach. The first
difference lies in the context. The consistent histories approach
describes
 individual systems. without making special reference to
measurement outcomes. Here we are only  interested in empirical
probabilities, as determined by measurements. The second difference
is more important: the consistent histories approach places no
interpretation on the decoherence functional or the objects
(\ref{PROB}), unless the former vanishes, in which case (\ref{PROB})
define a genuine probability measure. Here, in our search to
preserve the quantum logic at the level of measurements, we need to
find an interpretation of the mathematical object (\ref{decfun}) in
terms of observable quantities, namely the statistical behavior of
the sequence of relative frequencies $\nu_n$. Hence even if there
are many structural similarities between the present hypothesis and
consistent histories both the context and the physical
interpretation of the mathematical objects is conceptually distinct.

\subsection{Experimental distinction}

We mentioned two motivations for the hypothesis of non-converging
frequencies (a third more tentative one arising from the study of
frequency operators \cite{Fin63, Har68} can be found in Appendix C).
The main one was the preservation of the quantum logic structure of
sequential measurements. The decoherence functional for a history
depends on the sample sets only through the projectors. Hence the
statistical behavior of the relative frequencies that it
incorporates would be the same (modulo sampling and systematic
errors) in all different experiments that measure multi-time
probabilities. It would not depend, in particular, on whether the
experimental set-up is that of a sequential YES-NO experiment or of
a sequential measurement of position.

If this is true then it is very easy to distinguish the predictions
of this hypothesis from that of standard quantum theory. If the
quantum logic structure is preserved, probabilities for multi-time
samplings are not definable, but they are for single-time
measurements. In particular, the single-time marginals in a
multi-time measurement should always coincide with those of
single-time quantum theory. In a two-time measurement of an
observable $\hat{A} = \sum_i a_i \hat{P}_i$, the frequencies
$\nu_n(U_1, 0; U_2, t)$ do not converge for generic sample sets
$U_1$ and $U_2$, but the coarse-grained frequencies $\nu_n(\Omega,
0; U_2, t)$ should correspond to the single-time probabilities
$\sum_{i \in U_2} Tr\left( \hat{\rho} \hat{Q}_i\right)$, where
$\hat{Q}_i = e^{i \hat{H}t}\hat{P}_i e^{-i \hat{H}t}$. On the other
hand standard quantum theory predicts for the same probabilities
that
\begin{eqnarray}
p(\Omega, 0; U_2, t) = \sum_{j \in U_2} \sum_{i} Tr \left(
\hat{\rho}\hat{P}_i \hat{Q}_j \hat{P}_i \right). \label{condmarg}
\end{eqnarray}
The results are clearly different. For successive Stern-Gerlach
measurements, the first in the $x$ and the second in the $z$
direction of spin, the former hypothesis yields a probability
density $p_i = Tr (\hat{\rho} \hat{P}^z_i)$ with $\hat{P}^z_i$ the
spectral projectors of spin, while standard quantum theory yields a
constant probability density $p_i = \frac{1}{2}$. Since the
distinction arises at the level of the marginals, it is not
necessary to perform experiments with individual, distinguishable
runs, but it suffices to employ particle beams.

Note that the  local hidden variable theories of the type
considered in section 4.3 also satisfy equation (\ref{condmarg}),
 since their measurement outcomes are
described effectively by a stochastic process by virtue of
equations (\ref{derivation}, \ref{derivation2}).

\paragraph{ "Welcher-Weg experiments.}
One may contend that the behavior above can be excluded on the basis
of the so-called "welcher Weg experiments", in which a detector
placed immediately behind the holes of a two slit experiments
destroy the interference pattern. One may consider for example the
treatment of \cite{Scu}, which involves the interference of two
neutron beams. One places a micromaser cavity in the course of each
beam. The photons in the cavity interact with the neutrons' spin
degrees of freedom. If the field in the cavities is prepared      in
a number state, the interaction reveals unambiguously that the
neutron passed through one cavity or the other, and hence provides
information about the neutron's position. As a result the
interference pattern observed in a screen behind the detectors is
destroyed. Clearly, the intermediate measurement leads to different
results for the probability distribution on the screen and a
violation of (\ref{condmarg}).

There is however a flaw in this argument. Within standard quantum
theory, any measurement involves the separation between a quantum
system and a classical apparatus. This is not an easy distinction
to make, but it is necessary in order to specify the level at
which the reduction procedure is implemented. In the experiment
discussed above the electromagnetic field in the cavity is
described  by quantum mechanics. In the assumed splitting between
quantum and classical, it falls within the quantum domain.
 Hence, the quantum system in
consideration is not the neutron, but the combined system of neutron
and  electromagnetic field, which interact non-trivially through the
spin degrees of freedom. However, in absence of the micromaser the
quantum system is only the neutron. It is therefore not possible to
verify a violation of (\ref{condmarg}) by  comparing the
probabilities in these different experiments. Indeed, if the total
system of electromagnetic field and neutron is treated as quantum
mechanical,  the loss of interference is expected whether or not the
photon number has been measured in the microcavity. The distribution
of particle positions is obtained by the reduced density matrix of
the neutron  interacting with the electromagnetic field. This is
naturally expected to exhibit a loss of coherence, arising solely
from unitary evolution of the total system. To test equation
(\ref{condmarg}) one would have to compare the probability
distribution of neutron positions between an experiment that
includes a measurement of the photons in the cavity and one that
does not.

In any attempt to verify the validity of (\ref{condmarg}) one has to
compare experimental set-ups for which the split between quantum and
classical occurs at the same level. This is the case of the two-time
position measurement sketched in Fig.3. In that case one also has to
take into account all possible sources of error. In the original
Bohr-Einstein debate that led to the formulation of the
complementarity principle, the demonstration that "which-way"
information destroys the interference patter came essentially from
classical arguments about the inherent limitation in the precision
of  the first measurement \cite{Bohr}. Quantum theory was only
introduced, in order to place an upper limit in the measuring
accuracy. Bohr' argument  is therefore very different in character
than that of \cite{Scu}. It states essentially  that the uncertainty
at the level of the classical device affects  the quantum phases
randomly and thus leads to a destruction of the interference
pattern.

The same argument can be invoked in the present context  in relation
 to the assumption that measurements take place at a sharply
specified moment of time. This is not the case in a realistic
experiment. One may of course incorporate this uncertainty in the
error $\delta$ of an unsharp measurement. However, in multi-time
measurements there appears an additional source of randomness: the
presence of the first measuring device makes it more difficult to
specify the moment of time, at which the second measurement takes
place.

To see this, we may consider for example the thought-experiment of
Fig. 3. The detection of a particle can be assumed to take place
at a specific moment of time (the same for all runs of the
experiment) if the particle's momentum is sharply defined: $p_z$
is essentially treated as a classical variable. In a two-time
measurement however the value of $p_z$ changes randomly after the
interaction with the first measurement device, say by an amount of
$\delta p_z = d^{-1}$, for some parameter $d$ with the units of
length. This implies a fuzziness in the time that the particle
arrives in the second detector of the order of $\frac{t}{p_z d}$,
hence an additional uncertainty in the specification of the
particle's position of an order of $\frac{t}{md}$. The uncertainty
$d$ can be expected to be of the order of $\delta$, hence the
uncertainty is of the order of magnitude of $\frac{t}{m\delta}$

To see that the randomness induced by the uncertainty in the
specification of the measurement time is by itself responsible for
the loss of the interference patter, we consider the marginal
probability density induced by the POVM (\ref{Ralt}). For an
initial state corresponding to the set-up of a two slit experiment
\begin{eqnarray}
\psi_0(x) = \frac{1}{\sqrt{2} (2 \pi \sigma^2)^{1/4}} \left[ e^{
-\frac{(x - L/2)^2}{2 \sigma^2}} + e^{ -\frac{(x + L/2)^2}{2
\sigma^2}} \right],
\end{eqnarray}
where $\sigma$ is the width of the slit, $L$ the distance between
the slits and the mean momentum in the $x$-direction is for simplicity set
equal to zero \footnote{Note the $x$ direction is transverse to
the particle's direction of motion.} we obtain
\begin{eqnarray}
p(\Omega, 0; U, t) = \int_U dx \sqrt{\frac{m}{\pi t (\gamma +
\beta)} } \left[ e^{- \frac{(x - L/2)^2}{4(\gamma  + \beta)}} +
e^{- \frac{(x - L/2)^2}{4(\gamma  + \beta)}} \right. \nonumber \\
\left. + 2 e^{ - \frac{L^2}{4 \sigma^2}( 1 -
\frac{1}{\sigma^2(\gamma + \beta)})} e^{- \frac{m^2}{t^2(\beta +
\gamma)}x^2} \cos \left(\frac{mL}{t\sigma^2(\beta+\gamma)}x\right)
\right], \label{margx}
\end{eqnarray}
where $\beta = \frac{1}{2\delta^2} + \frac{2 m^2 \delta^2}{t^2}$
and $\gamma = \frac{1}{\sigma^2} + \frac{ m^2 \sigma^2}{t^2}$.
Hence, even the probability constructed from the consideration of
a two-time measurement in standard quantum mechanics exhibits
terms that describe  a distinguishable interference pattern. The
interference pattern is washed out only when an additional
consideration of the error due to the time uncertainty is taken
into account. The period of the interference pattern is $ c
\frac{t}{m L}$, where $c$ is a constant of order unity.  The
fuzziness  due to the uncertainty relation is of the order of
$\frac{t}{m \delta}$. For the two-slit experiment to make sense
the distance between the slits has to be much larger than the
resolution of the  measurement device--hence $L
>> \delta$. It follows that the interference pattern is hidden beneath the effect of the
time uncertainty.

It is important to stress the procedure we followed to obtain the
result above. Following Bohr's argumentation, the proof that the
interference patter disappears does not arise from the operational
rules of quantum theory about sequential measurements, but by
physical considerations that apply to our inherent inability to
establish with arbitrary precision the time a measurement takes
place. This effect cannot be obtained through formal manipulations
in standard quantum theory: at an operational level the formalism
allows one to talk only about measurements at sharp moment of
time\footnote{This problem is related to issue of constructing
time-of-arrival probabilities in quantum theory, and is explored
further in \cite{AnSav05}.}, while if we attempt to treat the
measuring apparatus as fully quantum mechanical we come face to face
with the measurement problem (is the reduction of the wave packet
instantaneous, if a physical process at all?).

The argument that proves that interferences are washed out in a
two-time position measurement works the same way in standard quantum
theory and within the non-converging frequency hypothesis. In the
latter case one also obtains an interference pattern with period of
the order $\frac{t}{m L}$, which is hidden beneath the effects of
the randomness due to time uncertainty. Even though the
right-hand-side is different from (\ref{margx}) the form of the
terms is the same and the difference only lies in the exact value of
the coefficients. Since this difference is drowned from the effects
of the time uncertainty, we conclude that it is very difficult (if
not impossible) to distinguish the predictions of quantum theory
from those of the non-converging frequency hypothesis (or of local
hidden variable theories for that matter) by means of equation
(\ref{condmarg}).

Hence the only way to distinguish between those theories would be
by directly measuring the frequencies $\nu_n(U_1,t_1;U_2,t_2)$ and
trying to establish whether they converge or not. In the Appendix
B we prove that this is in principle feasible, i.e. the suggested
failure of the frequencies to converge is much stronger than any
sources of error (such as the time uncertainty considered
earlier), and for this reason it is in principle detectable.

\section{Conclusions}
We conclude with a brief summary of the paper's results.

We studied the issue of constructing probabilities for sequential
measurements. In Section 3 we demonstrated that these probabilities
are highly contextual, namely they depend very strong on seemingly
trivial details of the apparatus (the parameter that determine its
resolution). We noted that this is a case of contextuality that does
not involve counterfactual reasoning: it may be determined in a
direct measurement set-up.

A key step for our analysis is the proof of a general theorem that
there is no way to reproduce the probability distribution for the
results of quantum theory from a stochastic process for the measured
system's degrees of freedom. We elaborated on this point in section
4, where we demonstrated that hidden variable theories can reproduce
the predictions of standard quantum theory only if they include
non-local interactions.

Finally, in section 5 we explored a rather unconventional
alternative that could allow the preservation of quantum logic in
sequential measurements: that probabilities are not defined, because
the corresponding frequencies do not converge. We demonstrated that
the predictions of this proposal can be unambiguously distinguished
from those of quantum theory.

\section{Acknowledgements}
I would like to thank N. Savvidou for many discussions and
suggestions in the course of this work. I have also benefited from
many discussions with D. Ghikas.  Research was supported partly by a
Marie Curie Reintegration Grant of the European Commission, the
Empirikion Foundation, and  by a Pythagoras II (EPEAEK) research
program.

\begin{appendix}

\section{Unsharp measurements and smeared characteristic functions}
In any measurement there are  systematic errors, uncertainties in
the specification of the initial state or the preparation of the
apparatus,fuzziness in the sampling of the results, dependence on
uncontrollable properties of the measurement device etc. For this
purpose it is necessary to consider the description of unsharp or
fuzzy measurements.

In unsharp measurements there will be outcomes for which we will
not be able to state unambiguously that
 "the system was found in the subset $U$ off the sample space
$\Omega$" or its negation. Such assertions can be made only with a
degree of confidence characterized by the relative size $\delta$ of
the measuring uncertainty. This can be implemented
 by substituting the characteristic functions that
represent the propositions about the measurement outcomes with
smeared characteristic functions $\chi_U^{\delta}$, which differ
from true characteristic functions on the scale of $\delta$. Given
the fact that a characteristic function for $\Omega = {\bf R}$ is
written as
\begin{eqnarray}
\chi_U(x) = \int_U dx' \delta(x-x'),
\end{eqnarray}
a smeared characteristic function may be written as
 \begin{eqnarray}
\chi_U^{\delta}(x) = \int_U dx' f_{\delta}(x-x'),
 \end{eqnarray}
 where $f_{\delta}$  is an one-parameter family of smooth
 functions converging weakly to the delta function as $\delta
 \rightarrow 0$. Any real-valued function that falls to zero rapidly
outside $U$, takes values close to unity well inside $U$ and
interpolates continuously between 1 and 0 in a region of size
$\delta$ around the boundary of $U$ is an adequate smeared
characteristic function. We may use for example a Gaussian family
\begin{eqnarray}
f_{\delta}(x) =  \frac{1}{2 \pi \delta^2} e^{-x^2/2\delta^2}.
\label{Gaussian}
\end{eqnarray}

 If the size of the sample set $U$ is $L$, then the difference of
the smeared characteristic function from a true characteristic
function is of the order of $O(\delta$. This difference may be
quantified by a norm in the space of functions--usually the $L^1$
norm-- of the difference $\chi_U - \chi_U^{\delta}$. Indeed, it is
easy to estimate that for the Gaussian smearing function
(\ref{Gaussian})

\begin{eqnarray}
 \int dx |\chi_U - \chi_U^{\delta}| < c \delta, \label{ineq1}
\end{eqnarray}
where $c$ is a positive number of order unity. This equation can
also be employed as a definition of a smeared characteristic
function. This implies that the relative difference between the
smeared and the genuine characteristic function is
\begin{eqnarray}
\epsilon = \frac{\int dx |\chi_U - \chi_U^{\delta}|}{\int_U dx } <
c \frac{\delta}{L}.
\end{eqnarray}

 It is not possible to obtain an equation analogous to
 (\ref{ineq1}) for the difference
$\chi_U - \chi_U^{\delta}$ weighted by the probability. However,if
we define the margin $M$ of the sample set $U$ as the region of
$\Omega$ in which $|\chi_U - \chi_U^{\delta}|$ is appreciably
larger than zero (say larger than a fixed small number $r <<1$),
we may estimate that
\begin{eqnarray}
\int \rho(x) |\chi_U - \chi_U^{\delta}| < \int_M  dx \rho(x) +
O(r),
\end{eqnarray}
hence for suitable choice of $r$ we can always find a constant $c$
of the order of unity such that
\begin{eqnarray}
\int \rho(x) |\chi_U - \chi_U^{\delta}| < c \int_M  dx \rho(x).
\end{eqnarray}
For general $\rho$ we cannot improve the above inequality. The
physically interesting case is one for which  $\rho$  varies at a
scale much larger than the size $\delta$ of the margin, otherwise
any probabilistic information would be completely lost beneath the
sampling error. In that case one may estimate that
\begin{eqnarray}
\int \rho(x) |\chi_U - \chi_U^{\delta}| < c' \frac{\delta}{L} p(U)
\end{eqnarray}
For the most general case the following estimation is relevant
\begin{eqnarray}
\int \rho(x) |\chi_U - \chi_U^{\delta}| < c' \frac{\delta}{R}
p(U),
\end{eqnarray}
 where $R$ is the size of the area of support of $\rho(x)
\chi)U(x)$.

\section{Generalization of  the results of section 4.2}

\paragraph{Unsharp measurements.} The result (\ref{derivation}) can be
reproduced even for unsharp measurements of a continuous pointer
function $X$. In that case one substitutes equations (\ref{cond1})
and (\ref{cond2}) with
\begin{eqnarray}
X[Q_t(x_0, Q_0)] = f(x_t(x_0, Q_0)) + O(\delta) \\
 p(U, t) = \int dx_0 , dQ_0 \, |\psi_0|^2(x_0) \, |\phi|^2(Q_0) \,
 \chi_U[X_t(x_0,Q_0)] = \nonumber \\
 \int dx_0 \, |\psi_0|^2(x_0) \,
 \chi^{\delta}_U[x^{sys}_t(x_0)] = \langle \psi_0|e^{i \hat{H_0}t}\hat{F}^{\delta}_U e^{-i \hat{H_0}t}|\psi)
 \rangle,
\end{eqnarray}
where now $\hat{F}^{\delta}$ is a POVM for the variable $f$
characterized by a parameter $\delta$, which incorporates the
effects of the interaction with the measurement device. The proof
follows the same steps as the discrete variable case, the only
difference being that we substitute the characteristic functions
with smeared ones. Since for a family of characteristic functions
$\chi^{\delta}$ labeled by $\delta$
\begin{eqnarray}
\chi^{\delta}_{U_1} \chi_{U_2}^{\delta} = \chi^{\delta}(U_1 \cap
U_2) + O(\delta), \label{chis}
\end{eqnarray}
it is easy to conclude that if the locality condition
(\ref{locality}) the probabilities $p^{2}$ coincide up to an error
of order $O(\delta)$ with those obtained by a stochastic process
constructed from the degrees of freedom of the system by itself.
 Hence for
samplings of size much larger than $\delta$ the probabilities do not
depend on properties of the measurement device, something that
contrasts the results of standard quantum theory like Eq.
(\ref{Rn}).

\paragraph{Markov process.} The same arguments may be applied for
non-deterministic hidden variable theories that reproduce the
predictions of quantum theory through a Markov process. We denote by
$g(x, Q, t| x', Q', t')$ the propagator corresponding to the
interacting dynamics between system and apparatus, by $h(Q,t|Q',t')$
the one corresponding to the self-dynamics of the apparatus and by
$g^{sys}(x,t|x',t')$ the one corresponding to the
 self-dynamics of the system {\em in absence of apparatus}.

Assuming an initially factorized state the conditions that the
stochastic process reproduces the operational predictions for
single-time quantum theory for a discrete pointer $X$ are
\begin{eqnarray}
X(Q_t) &=& f(x_t), \\
 p(U,t) &=& \int dx_0 \, dQ_0 \, dx_t \, dQ_t \, |\psi_0|^2(x_0) \, |\phi|^2(Q_0) \nonumber \\
 &\times& g(x_0, Q_0,
0| x_{t}, Q_{t}, t) \, \chi_U[X(Q_t))] \nonumber \\
&=& \int dx_0 \, dx_t \, |\psi_0|^2(x_0) g^{sys}(x_0,0|x_t,t)
\chi_U(f(x_{t_t}))
\end{eqnarray}
where $t$ denotes the time that the measurement has been completed.

In a two-time measurement the total propagator for the degrees of
freedom of the measured system and the two measuring devices
factorized as $g(x_0, Q_0, 0; x^1_t, Q^1_t, t)$ $\times
g(Q^2,0|Q^2_t, t)$ for all times prior to the first measurement, as
the two devices are assumed non-interacting. The key assumption,
analogue to the locality postulate (\ref{locality}) in deterministic
systems is that the propagator for the total system factorized {\em
after} the first measurement as
\begin{eqnarray}
g(x_{t}, Q^2_t, t| x_{t'}, Q^2_{t'},t') h(Q_t, t| Q_{t'}, t'),
\hspace{1cm} t,t' > t_1, \label{locality2}
\end{eqnarray}
which states that the particle's stochastic evolution is not
affected by the degrees of freedom of the first apparatus after the
interaction has been completed.

With the assumptions above, it is easy to show following steps
analogous to those of (\ref{derivation}) that the two-time
probability
\begin{eqnarray}
p(U_1, t_1; U_2, t_2) = \int dx_0 \, dQ^1_0 \, dQ^2_0 \;  dx_{t_1}
\, dQ^1_{t_1} \, dQ^2_{t_1} \; dx_{t_2} \, dQ^1_{t_2} \, dQ^2_{t_2}
\nonumber
\\
\times |\psi_0|^2(x_0) \, |\phi_1|^2(Q^1_0) \, |\phi_2|^2(Q^2_0)
g(x_0, Q^1_0,0|x_{t_1}, Q^1_{t_1}, t_1) \,
h(Q^2_0,0|Q^2_{t_1}, t_1)  \nonumber \\
\times  \chi_{U_1}[X_{t_1}] g(x_{t_1}, Q^2_{t_1}, t_1|x_{t_2},
Q^2_{t_2}, t_2) \, h(Q^1_{t_1}, t_1|Q^1_{t_2}, t_2) \,
\chi_{U_2}[X_{t_2}]
\end{eqnarray}
equals
\begin{eqnarray}
p(U_1, t_1; U_2, t_2) = \int dx_0 dx_{t_1} dx_{t_2} \;
|\psi_0|^2(x_0) \;
g^{sys}(x_0,0|x_{t_1}, t_1)  \nonumber \\
\times \chi_{U_1}(f_{t_1}) \, g^{sys}(x_{t_1}, t_1|x_{t_2}, t_2) \,
\chi_{U_2}(f_{t_2}). \label{derivation2}
\end{eqnarray}

Hence probabilities are again described by a stochastic processes
for the measured system's degrees of freedom, in conflict with the
predictions of quantum theory. To reproduce the predictions of
quantum theory with a Markov process, one would need to assume a
violation of Eq. (\ref{locality}).
\section{Frequency operators}

 In some interpretations of quantum theory, it is often asserted that the Born's rule
from probabilities can be obtained from a weaker postulate, namely
that if an observable $\hat{A}$ is measured on a system in one of
its eigenstates, the outcome is the corresponding eigenstate. The
idea is to construct a Hilbert space for the statistical ensemble
${\cal H}_{ens}$ as (ideally an infinite) tensor product
$\otimes_n{\cal H}_n$ of the Hilbert space ${\cal H}$ of the single
system \cite{Fin63, Har68} (see also \cite{Mit04} and references
therein). Assuming a projection operator $\hat{P}$ corresponding to
a state $|i \rangle$ of ${\cal H}$, we may construct a PVM
corresponding to the different values of the frequencies $f$ for the
event corresponding to $\hat{P}$ in the statistical ensemble. For a
finite ensemble consisting of $N$ copies this PVM reads
\begin{eqnarray}
\hat{\Pi}_{\hat{P}}(f = n/N) = \sum_{k1+k_2+ \ldots + K_n = n }
\hat{P}_{k_1} \otimes \ldots \otimes \hat{P}_{k_n}, \label{freqPVM}
\end{eqnarray}
where $k_i = 1$ corresponds to $\hat{P}_{k_i} = \hat{P}$ and $k_i =
1$ corresponds to $\hat{P}_{k_i} = \hat{1} - \hat{P}$.

One then may (attempt to) prove that the vectors $\otimes_n |\psi
\rangle_n$ are eigenstates of the frequency operators
$\hat{F}_{\hat{P}} = \sum_{n = 0}^N \frac{n}{N}
\hat{\Pi}_{\hat{P}}(f= n/N)$ at the limit $N \rightarrow \infty$
with eigenvalues coinciding to the standard probabilities $|\langle
i|\psi \rangle|^2$.

The underlying concept in this approach is that the Born rule may be
derived {\em solely} from the projective geometry of the Hilbert
space by making reference to the ensemble as an individual quantum
system. This approach faces some problems in its mathematical
implementation \cite{CS05}, but it is  interesting to see whether it
can be applied to sequential measurements.

The key obstacle is that one cannot assign projection operators
corresponding to the outcomes of sequential measurements. Even for
ideal measurements the best we can do is to construct a POVM like
(\ref{POVM2}), in which two successive readings $i$ and $j$ of the
variable $\hat{x}$ will correspond to the positive operators
$\hat{K}_{ij} = \hat{P}_i e^{i \hat{H}t} \hat{P}_j e^{- i \hat{H}t}
\hat{P}_i$. The analogue of the PVM (\ref{freqPVM}) would therefore
be a POVM $\hat{\Pi}_{\hat{K}}$, in which $\hat{K}_{ij}$ would be
inserted in place of the projector $\hat{P}$. It is easy to verify
that the failure of the idempotency condition $\hat{K}_{ij}^2 =
\hat{K}_{ij}$ implies that
\begin{eqnarray}
\hat{\Pi}_{\hat{K}}(f_{ij} = n/N) \hat{\Pi}_{\hat{K}}(f_{ij}= n'/N)
\neq 0 \; \; \mbox{if} \; n \neq n',           \label{n0}
\end{eqnarray}
and that this property persists even at the limit $N \rightarrow
\infty$. These POVMs cannot  distinguish between different values of
the frequency in the ensemble.  It is, therefore, not possible to
obtain probabilities solely from the geometry of Hilbert space,
because the positive operators $\hat{\Pi}_{\hat{P}}(f = n/N)$ cannot
be associated with specific values of frequency in a way that
respects the projective character of Hilbert space geometry.

Alternatively one could define a POVM corresponding to frequencies
$f_1 = n_1/N$ for the outcome $i$ of the first measurement and $f_2
= n_2/N$ for the second
\begin{eqnarray}
\Pi(f_1= n_1/N,0; f_2 = n_2/N, t) = \hspace{5cm}\nonumber \\
\hat{\Pi}(f_1 = n_1/N) [\otimes_n e^{i \hat{H}t} ] \hat{\Pi}(f_2 =
n_2/N) [\otimes_n e^{-i \hat{H}t}] \hat{\Pi}(f_1 = n_1/N).
\end{eqnarray}
This POVM is different from the one obtained from the modification
of (\ref{freqPVM}), because different fine-grained alternatives are
used in its construction. It is still subject to Eq. (\ref{n0}) and
cannot distinguish between different values of frequency.

The results above imply that  the programme of defining
probabilities through frequencies (without {\em a priori} assuming
Born's rule)  cannot account for sequential measurements. The only
way to salvage it, is to take Eq. (\ref{n0}) at face value and
assume that different values of the frequency cannot be
distinguished
 in sequential measurements, implying in effect that multi-time probabilities
 are ill-defined--or they do not converge.

\section{Distinguishability of non-converging
frequencies}

We consider a two-time measurement of position as described in
section 3.2. We assume that a beam of free particles with mass
$m$ is prepared in a state $\psi_0$, centered around $x = 0$, with
zero mean momentum $p_x$ and a spread $L$ in the $x$ direction. We
consider for simplicity only two samplings at each moment of time,
corresponding to the sets $U_+ = [0,\infty)$ and $U_- = (-
\infty,0]$. The corresponding projectors are $\hat{P}_+$ and
$\hat{P}_-$. We then consider the candidate probability that the
particle is detected in $U_+$ at time $t_1$ and in $U_2$ at time
$t_2$,
\begin{eqnarray}
 p_{++} = \langle \psi_0| \hat{P}_+ e^{i\hat{H}t} \hat{P}_+
e^{-i\hat{H}t} \hat{P}_+|\psi_0 \rangle,
\end{eqnarray}
 while the obstruction to additivity equals
\begin{eqnarray}
 b = 2 Re \langle
\psi_0| \hat{P}_+ e^{i\hat{H}t} \hat{P}_+ e^{-i\hat{H}t}
\hat{P}_-|\psi_ \rangle.
\end{eqnarray}
\begin{figure}[h]
\includegraphics[height=5cm]{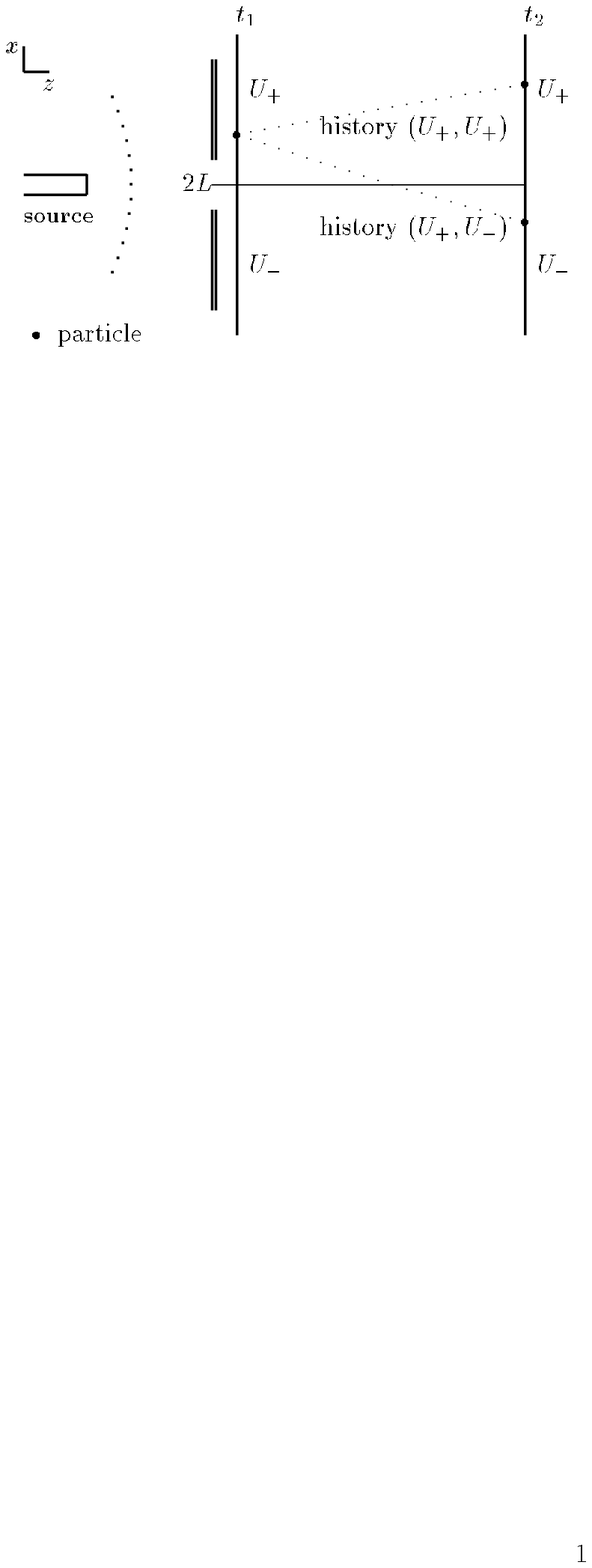} \caption{ \scriptsize The particles pass through a slit of width
$2L$ and are registered in two successive screens. }
\end{figure}

The details of the wave-function's shape do not significantly
affect the result. For calculational convenience we consider
\begin{eqnarray}
\psi_0(x) = \frac{1}{\sqrt{2L}} \chi_{[-L,L]}(x),
\end{eqnarray}
 where $
\chi_{[-L,L]}$ is the characteristic function of the set $[-L,L]$
-- corresponding for instance to a slit of width $2L$ placed
immediately before the first detector. We then obtain
\begin{eqnarray}
p_{++} = \frac{1}{\pi} \int_0^1 dz Si[z^2/r] \nonumber \\
b = \frac{r}{\pi} \int_0^1 \frac{1 - \cos(z^2/r)}{z^2},
\end{eqnarray}
 where
 $r = \frac{t}{mL^2}$, the dimensionless time-scale; $Si$
stands for the sin-integral function. In Fig. 6 we plot the ratio
$b/p_{++}$ as a function of $r$: it starts from $0$ at $r = 0$ (in
which case we have a single-time measurement), it increases
rapidly, and for $r \sim 1$ reaches the asymptotic value $1/2$. In
other words, the assumed non-convergence of probabilities is
manifested very strongly even for the highly coarse-grained sample
sets $U_+$ and $U_-$.

\begin{figure}
 \includegraphics[height=3.5cm]{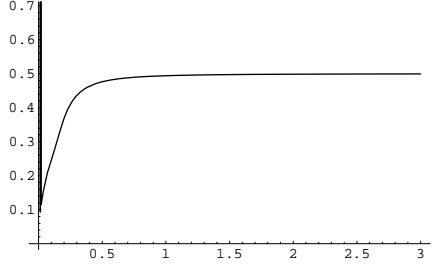}
 \caption{ \scriptsize A plot of the ratio $b/p_{++}$ versus dimensionless time $r = \frac{t}{m L^2}$ for the system of Fig. 2.
 This ratio measures the failure of additivity in
 the natural probability assignment and  estimates the relative size of the area of non-convergence
for event frequencies.} \label{qbp}
 \end{figure}

In any statistical sample there exists an extended region of
convergence for relative frequencies due to sampling errors or
systematic uncertainties, or due to the finite number of
experimental runs. These errors can be accounted for by positive
operators that approximate projectors within an order of the
error. If $d$ is the size of the error, we need to employ
operators of the form
 $ \int
\chi_{U}^d(x)|x \rangle \langle x|$, defined by the smeared
characteristic functions  $\chi_{U}$. The related error
 equals $|Tr \hat{\rho}
(\hat{P}_U - \hat{P}_U^2)| \sim \frac{d}{R} Tr(\hat{\rho}
\hat{P}_U)$, where $R$ is the size of the support of $\rho(x,x)
\chi_U(x)$. For the configuration of Fig. 4,  $R \sim L$ and  $d$
is  the width of the particle's trace, so the error is at most of
the order of $d/L$.

The operators describing the sampling of measurement outcomes have
also to incorporate the indeterminacy in measurement time--see the
discussion in section V.5 . One should therefore employ a {\em
time-averaged} projector, on an interval of width $\tau$ around
the moment of time $t$. This corresponds to the proposition that
the measurement took place at any time within the interval
$[t-\tau/2, t + \tau/2]$,
\begin{eqnarray}
\hat{\Pi}_U = \frac{1}{\tau}\int_{t- \tau/2}^{t+ \tau/2} ds
e^{i\hat{H} s}\hat{P}_Ue^{-i\hat{H} s}.
\end{eqnarray}
 To leading
order in $\tau$, the spread of $\hat{\Pi}_U$ is $\epsilon =
\frac{\tau}{m d}$. The indeterminacy in time $\tau$ is related to
the interaction of the system with the first sheet and should be of
the order $\frac{\delta p _z}{p_z}t$, where $\delta p_z$ is the
momentum transfer in the $z$ direction as the particle crossed the
sheet. Hence we estimate that $\epsilon \sim \frac{\delta p_z}{m p_z
d}t $. The uncertainty relation suggests that $\delta p_z$ is of the
order of $1/d$, so $\epsilon \sim\frac{t}{m^2 v_z d^2}$, where $v_z$
is the mean velocity of the particles in the $z$ direction. Since
the non-additivity of probabilities is manifested for $r \sim 1 $,
we may substitute the corresponding value of $t$ to obtain $
\epsilon \sim \frac{L}{m d^2 v_z}$. Hence the total uncertainty in
the measurement is of the order of $c_1 \frac{d}{L} + c_2 \frac{L}{m
d^2 v_z}$, with $c_1, c_2$ numbers of order unity.  For realistic
values  of $L = 1 \, cm$, $d = 10^{-2}\, cm$, $v_z = 10^4 \, m/s$,
the error due to time indeterminacy is of the order of $10^{-4}$,
much smaller than the ratio $d/L \sim 10^{-2}$ of relative error in
position sampling. It follows that the non-convergence of
probabilities--if present-- is more pronounced than the measurement
uncertainties and can be in principle detected.

\end{appendix}

\begin{thebibliography}{}

\bibitem{Bell64} J. S. Bell,
\newblock {\em On the Einstein-Podolsky-Rosen Paradox,} Physics 1, 195 (1964).

\bibitem{Wig76} E. Wigner, {\em On Hidden Variables and Quantum
MEchanical Probabilities}, in Mathematical Physics and Applied
Mathematics, Vol.1, 33,  Dordrecht (1976).

 \bibitem{KoSp67} S. Kochen and R. P. Specker, {\em The Problem of Hidden Variables in Quantum Mechanics},
 J. Math. Mech. 17, 59 (1967).



\bibitem{Ana01a} C. Anastopoulos, {\em Quantum Theory without Hilbert Spaces},
Found. Phys. 31, 1545 (2001).

\bibitem{Ana03a} C. Anastopoulos, {\em Quantum Processes on Phase Space}, Ann. Phys. 303, 275
(2003).

\bibitem{Sor9497} R. D. Sorkin,  {\em Quantum Mechanics as  Quantum Measure
  Theory},  Mod. Phys. Lett.  A9, 3119 (1994);  {\em Quantum Measure Theory and its
  Interpretation},
 in {\em  Quantum Classical Correspondence}, edited by D.H. Feng and B. L. Hu, (International Press, Cambridge MA, 1997).

\bibitem{Nel66} E. Nelson, {\em Derivation of Schr\"odinger's
Equation from Newtonian Mechanics},  Phys. Rev. 150, 1079 (1966).

\bibitem{Nel85} E. Nelson, {\em Quantum Fluctuations}, (Princeton
University Press, Princeton, 1985).


\bibitem{Gri84} R. Griffiths,
 {\em Consistent Histories and the  Interpretation
of Quantum Mechanics},
   J. Stat. Phys.  36, 219 (1984).

 \bibitem {Omn8894} R. Omn\`es,
\newblock  {\em Logical Reformulation of Quantum Mechanics:
I Foundations},
\newblock   J. Stat. Phys.   53, 893 (1988);
\newblock  {\em The Interpretation of Quantum Mechanics},
\newblock  (Princeton University Press, Princeton, 1994);
\newblock  {\em Consistent Interpretations of Quantum Mechanics},
\newblock  Rev. Mod. Phys. 64, 339 (1992).

  \bibitem {GeHa9093} M. Gell-Mann and J. B. Hartle,
 \newblock  {\em Quantum mechanics in the Light of Quantum Cosmology},
\newblock  in {\em Complexity, Entropy and the Physics of Information},
edited by W.\ Zurek,
\newblock    (Addison Wesley, Reading, 1990);
\newblock  {\em Classical Equations for Quantum Systems},
\newblock  Phys. Rev. D   47, 3345 (1993).

\bibitem{Har93a} J. B. Hartle,
\newblock {\em Spacetime Quantum Mechanics and the Quantum Mechanics of Spacetime},
\newblock in   Proceedings on the 1992 Les Houches School,
Gravitation and Quantisation, 1993.

\bibitem{ABL64} Y. Aharonov, P. G. Bergmann and J. L. Lebowitz, {\em
Time Symmetry in the Quantum Process of Measurement}, Phys. Rev.
134, B1410 (1964).



\bibitem{DaLe71} E. B. Davies and J. T. Lewis, {\em An Operational Approach to Quantum
Probability}, Comm. Math. Phys. 17, 3 (1971).

\bibitem{Davies} E. B. Davies, {\em Quantum Theory of Open Systems}, (Academic Press, London, 1976).

\bibitem{AAD85}   D. Z. Albert, Y. Aharonov and S. D' Amato, {\em Multiple-time properties of quantum-mechanical
systems}, Phys. Rev. D32, 1975 (1985).

\bibitem{Cave86} C. M. Caves, {\em Quantum Mechanics of Measurements Distributed in Time. A Path-integral
Formulation}, Phys. Rev. D33, 1643 (1986).


\bibitem{MiSu77} B. Misra and E. C. G. Sudarshan, {\em The Zeno's Paradox in Quantum Theory}, J. Math. Phys.
18, 657 (1977).

\bibitem{BCL90} P.Busch, G. Cassinelli and P. Lahti, {\em On the
Quantum Theory of Sequential Measurements}, Found. Phys. 20, 757
(1990).


\bibitem{Hal93} J. J. Halliwell, {\em Quantum-mechanical Histories and the Uncertainty Principle: Information-theoretic
Inequalities}, Phys. Rev. D 48, 2739 (1993).



\bibitem{SN01} S. Gudder and G. Nagy, {\em Sequential quantum
measurements}, J. Math. Phys. 42, 5212 (2001).

\bibitem{Hol01} A.S. Holevo, {\em Statistical Structure in Quantum
Theory}, (Springer, New York, 2001).

 \bibitem{Ana04a} C. Anastopoulos, {\em On the relation between quantum mechanical probabilities and event
 frequencies}, Ann. Phys. 313, 368 (2004).

 \bibitem{vNeu} J. von Neumann,
 \newblock {\em The Mathematical Foundations of Quantum Mechanics}, (Princeton University Press, Princeton 1996).

\bibitem{BLM96} P. Busch, P. Lahti and Peter Mittelstaedt, {\em The Quantum Theory of
Measurement}, (Springer Verlag, Berlin, 1996).

\bibitem{BL96} P. Busch and P. Lahti, {\em The Standard Model of Quantum Measurement Theory: History and
applications}, Found. Phys. 26, 875 (1996).


\bibitem{BG00} A. Bassi and G. Ghirardi, {\em A General Argument
against the Universal Validity of the Superposition Principle},
Phys. Lett. A275, 373 (2000).

\bibitem{Schl04} M. Schlosshauer, {\em Decoherence, the Measurement Problem, and Interpretations of Quantum
Mechanics}, Rev. Mod. Phys. 76, 1267 (2004).

\bibitem{Adl01} S. L. Adler, {\em Why Decoherence has not Solved the Measurement Problem: A Response to P. W.
Anderson}, quant-ph/0112095.

\bibitem{Das05} T. Dass, {\em Measurements and Decoherence},
quant-ph/0505070.


\bibitem{AAD85a} D. Z. Albert, Y. Aharonov and S. D' Amato, {\em Curious New Statistical Prediction of Quantum
Mechanics}, Phys. Rev. Lett. 54, 5 (1985). See also: J. Bub and H.
Brown, {\em Curious Properties of Quantum Ensembles Which Have Been
Both Preselected and Post-Selected}; D. Z. Albert, Y. Aharonov and
S. D' Amato, {\em Curious Properties of Quantum Ensembles Which Have
Been Both Preselected and Post-Selected}, Phys. Rev. Lett. 56, 2427
(1986).

\bibitem{Kent} A. Kent, {\em Consistent Sets Yield Contrary Inferences in Quantum
Theory},  Phys. Rev. Lett. 78, 2874 (1997). See also: R. B.
griffiths and J. B. Hartle, {\em Comment on "Consistent Sets Yield
Contrary Inferences in Quantum Theory"}, Phys. Rev. Lett. 81, 1981
(1998).



\bibitem{BH} See for example D. Bohm and B. J. Hiley, {\em The Undivided Universe},
 (Routledge, 1995).


 \bibitem{Neu00} A. Neumaier, {\em Bohmian Mechanics Contradicts
 Quantum Mechanics}, quant-ph/0001011.

 \bibitem{FPSW02} L. Feligioni, O. Panella, Y. N. Srivastava, A.
 Widom, {\em Two-time Correlation Functions: Bohm Theory and
 Conventional Quantum Mechanics}, Eur. Phys. J. B 48, 233 (2005).

 \bibitem{Har04} J. B. Hartle, {\em Bohmian Histories and
 Decoherent Histories}, Phys. Rev. A69, 042111 (2004).


\bibitem{Bohmmeas} D. Bohm, {\em A Suggested Interpretation of the
Quantum Theory in Terms of Hidden Variables II}, Phys. Rev. 85, 180
(1952); D. Bohm and B. J. Hiley, {\em  Measurement Understood
Through the Quantum Potential Approach}, Found. Phys. 14, 255
(1984); J. S. Bell, {\em Quantum Mechanics for Cosmologists},
Quantum Gravity 2; A Second Oxford Symposium, Edited by C. J. Isham,
R. Penrose and D. W. Sciama, ( Clarendon Press, Oxford, England,
1981); D Duerr, S Goldstein, N Zanghi, {\em Quantum Equilibrium and
the Origin of Absolute Uncertainty}, J. Stat. Phys. 67, 843 (1992).


\bibitem{Stochmeas} P Blanchard, S Golin, M Serva, {\em Repeated Measurements in Stochastic
Mechanics}, Phys. Rev. D 34, 3732 (1986); S. Goldstein, {\em
Stochastic Mechanics and Quantum Theory}, J. Stat. Phys. 47, 645
(1987); G. Peruzzi and A. Rimini, {\em Quantum Measurement in a
Family of Hidden-Variable Theories}, Found. Phys. Lett. 9, 505
(1996).



\bibitem{Pop94} S. Popescu, {\em Bell's Inequalities and Density Matrices: Revealing "Hidden"
Nonlocality}, Phys. Rev. Lett. 74, 2619 (1995).

\bibitem{P3H98} Marek Zukowski, Ryszard Horodecki, Michal Horodecki and  Pawel Horodecki,
{\em Generalized Quantum Measurements and Local Realism}, Phys. Rev.
A 58, 1694 (1998).

\bibitem{LeGa86} A. J. Leggett and A. Garg, {\em Quantum Mechanics versus Macroscopic Realism: Is the Flux There when Nobody
Looks?}, Phys. Rev. Lett. 54, 857 (1985).

\bibitem{PaMa93} J. P. Paz and G\"unter Mahler, {\em Proposed Test
for Temporal Bell Inequalities}, Phys. Rev. Lett. 71, 3235 (1993).

\bibitem{CCO99} T. Calarco, M. Cini, R. Onofrio, {\em Are Violations to Temporal Bell Inequalities There when Somebody
Looks?}, Europhys. Lett, 47, 407-413 (1999).


\bibitem{surreal} B.G. Englert, M. O. Scully, G. S\"ussmann and H. Walther, {\em  Surrealistic Bohm trajectories},
Z. Naturforch. 48a, 1261 (1993); Y. Aharonov and L. Vaidman, {\em
About Position Measurements Which Do Not Show the Bohmian Particle
Position}, in "Bohmian Mechanics and Quantum Theory: An Appraisal,"
edited by J.T. Cushing, A. Fine, and S. Goldstein (Kluwer,
Dordrecht, 1996); B. J. Hiley, R. E. Callaghan and O. Maroney, {\em
    Quantum trajectories, real, surreal or an approximation to a deeper
    process?}, quant-ph/0010020.

 \bibitem{hooft} G. 't Hooft, {\em Determinism in Free Bosons},  Int. J. Theor. Phys. 42, 355
 (2003).




\bibitem{Fin63} D. Finkelstein, {\em  The Logic of Quantum Physics}, Trans. N.Y. Acad. Sci. 25, 621
(1963).

\bibitem{Har68} J. B. Hartle, {\em Quantum Mechanics of Individual Systems}, Am. J. Phys. 36, 704 (1968).

\bibitem{Mit04} P. Mittelstaedt, {\em The Interpretation of Quantum Mechanics and the Measurement
Process}, (Cambridge University Press, 2004).

\bibitem{CS05} C. M. Caves and R. Schack, {\em Properties of the
Frequency Operator do not Imply the Quantum Probability Postulate},
Ann. Phys. 315, 123 (2005).


\bibitem{Scu} M. O. Scully and H. Walther, {\em Quantum Optical
Tests of Observation and Complementarity in Quantum Mechanics},
Phys. Rev. A39, 5229 (1989). See also: M. O. Scully and K. Druhl,
{\em Quantum Eraser: A Proposed Photon Correlation Experiment
Concerning Observation and "Delayed Choice" in Quantum Mechanics},
Phys. Rev. A 25, 2208 (1982); S. M. Tan and D. F. Walls, {\em Loss
of Coherence in interferometry}, Phys. Rev. A 47, 4663 (1993); B. G.
Englert, {\em Fringe Visibility and Which-Way Information: An
Inequality}, Phys. Rev. Lett. 77, 2154 (1996).


\bibitem{Bohr} See for example: M. Jammer, {\em The Philosophy of
Quantum Mechanics}, (Wiley, New York, 1974).


 \bibitem{AnSav05} C. Anastopoulos and N. Savvidou, {\em
 Time-of-Arrival Probabilities and quantum measurement}, quant-ph/0509020.
\end{thebibliography}
\end{document}